\documentclass[aos,MSNbibl,seceqn,dvips]{arximspdf}
\usepackage{graphicx}
\doi{10.1214/14-AOS1226} 
\volume{42}
\issue{3}
\pubyear{2014}
\firstpage{1203}
\lastpage{1231}

\makeatletter
\newcommand{\SGN}{\mathrm{SGN}}
\newcommand{\SD}{\mathrm{SD}}
\newcommand{\sign}{\operatorname{sign}}
\renewcommand{\span}{\operatorname{span}}
\newcommand{\Var}{\operatorname{Var}}
\newcommand{\trace}{\operatorname{trace}}
\newcommand{\rrVert}{\Vert}
\newcommand{\rrvert}{\vert}
\newcommand{\llVert}{\Vert}
\newcommand{\llvert}{\vert}

\newtheorem{teo}{Theorem}[section]
\newproclaim{fact}{Fact}[section]
\newproclaim{ass}{Assumption}
\newproclaim{rem}{Remark}
\newtheorem{lemma}[fact]{Lemma}
\newtheorem{proposition}[fact]{Proposition}
\makeatother

\begin{document}
\begin{frontmatter}

\title{The spatial distribution in infinite dimensional spaces and
related quantiles and depths}
\runtitle{Spatial distributions, quantiles and depth}

\begin{aug}
\author{\fnms{Anirvan}~\snm{Chakraborty}\corref{}\ead[label=e1]{vanchak@gmail.com}\thanksref{t1}}
\and
\author{\fnms{Probal}~\snm{Chaudhuri}\ead[label=e2]{probal@isical.ac.in}}
\runauthor{A. Chakraborty and P. Chaudhuri}
\affiliation{Indian Statistical Institute}
\address{Theoretical Statistics and Mathematics Unit\\
Indian Statistical Institute\\
203, B.T. Road\\
Kolkata 700108\\
India\\
\printead{e1}\\
\phantom{E-mail: }\printead*{e2}}
\end{aug}
\thankstext{t1}{Supported in part by the SPM Fellowship of the Council
of Scientific and Industrial Research, Government of India.}

\received{\smonth{11} \syear{2013}}
\revised{\smonth{4} \syear{2014}}

%
\begin{abstract}
The spatial distribution has been widely used to develop various
nonparametric procedures for finite dimensional multivariate data. In
this paper, we investigate the concept of spatial distribution for data
in infinite dimensional Banach spaces. Many technical difficulties are
encountered in such spaces that are primarily due to the noncompactness
of the closed unit ball. In this work, we prove some Glivenko--Cantelli
and Donsker-type results for the empirical spatial distribution process
in infinite dimensional spaces. The spatial quantiles in such spaces
can be obtained by inverting the spatial distribution function. A
Bahadur-type asymptotic linear representation and the associated weak
convergence results for the sample spatial quantiles in infinite
dimensional spaces are derived. A study of the asymptotic efficiency of
the sample spatial median relative to the sample mean is carried out
for some standard probability distributions in function spaces. The
spatial distribution can be used to define the
spatial depth in infinite dimensional Banach spaces, and we study the
asymptotic properties of the empirical spatial
depth in such spaces. We also demonstrate the spatial quantiles and the
spatial depth using some real and simulated functional data.
\end{abstract}

%
\pdfkeywords{62G05, 60B12, 60G12, Asymptotic relative efficiency, Bahadur representation,
DD-plot, Donsker property, Gateaux derivative, Glivenko-Cantelli property,
Karhunen-Loeve expansion, smooth Banach space}
\end{frontmatter}

\section{Introduction}\label{intro}
The univariate median and other quantiles have been extended in
a number of ways for multivariate data and distributions in finite
dimensional spaces (see, e.g., \cite{DG92,Liu90,Oja83}
and \cite{Smal90}). In particular, the spatial median, the spatial
quantiles and the associated spatial distribution function in finite
dimensional Euclidean spaces have been extensively studied (see, e.g.,
\cite{Brow83,Chau96,Kolt97,MOT97} and \cite
{Serf02}). Nowadays, we often come across data, which are curves or
functions and can be modeled as random observations from probability
distributions in infinite dimensional spaces. The ECG curves of
patients observed over a period of time, the spectrometry curves
recorded for a range of wavelengths, the annual temperature curves of
different places, etc., are examples of such data. Many of the function
spaces, where such data lie, are infinite dimensional Banach spaces.
However, many of the well-known multivariate medians like the
simplicial depth median (see \cite{Liu90}), and the simplicial volume
median (see \cite{Oja83}) do not have meaningful extensions into such
spaces. On the other hand, the spatial median as well as the spatial
quantiles extend easily into infinite dimensional Banach spaces (see
\cite{Chau96,Kemp87} and \cite{Vala84}). The author of \cite
{Gerv08} proposed functional principal components using the sample
spatial median and used those to analyze a data involving the movements
of the lips. The authors of \cite{CCZ13} considered an updation based
estimator of the spatial median, and used it to compute the profile of
a typical television audience in France throughout a single day. The
spatial median has also been used in~\cite{CG12} to calculate the
median profile for the electricity load data in France. Recently, the
authors of \cite{FPL12} studied some direction-based quantiles for
probability distributions in infinite dimensional Hilbert spaces. These
quantiles are defined for unit direction vectors in
such spaces, and they extend the finite dimensional quantiles
considered in \cite{KM12}. The principle quantile directions derived
from these quantiles were used in \cite{FPL12} to detect outliers in a
dataset of annual age-specific mortality rates of French males between
the years 1899 and 2005.

The purpose of this article is to investigate the spatial
distribution in infinite dimensional Banach spaces, and study their
properties along with the spatial quantiles and the spatial depth.
There are several mathematical difficulties in dealing with the
probability distributions in such spaces. These are primarily due to
the noncompactness of the closed unit ball in such spaces. In
Section~\ref{sec2}, we prove some Glivenko--Cantelli and Donsker-type
results for the empirical spatial distribution process arising from
data lying in infinite dimensional spaces. In Section~\ref{sec4}, we
investigate the spatial quantiles in infinite dimensional spaces. A
Bahadur-type linear representation of the sample spatial quantiles and
their asymptotic Gaussianity are derived. We also study the asymptotic
efficiency of the sample spatial median relative to the sample mean for
some well-known probability distributions in function spaces. In
Section~\ref{sec3}, we investigate the spatial depth and its asymptotic
properties in infinite dimensional spaces. We also demonstrate how
exploratory data analytic tools like the depth--depth plot (DD-plot)
(see \cite{LPS99}) can be developed for data in infinite dimensional
spaces using the spatial depth. The proofs of the theorems are given in
the \hyperref[sec5]{Appendix}.

\section{The spatial distribution and the associated empirical processes in Banach spaces}\label{sec2}
For probability distributions in $\mathbb{R}^{d}$, the spatial
distribution is a
special case\vadjust{\goodbreak} of the $M$-distribution function, which was studied in
details in~\cite{Kolt97}.
Consider the map $f\dvtx  \mathbb{R}^{d} \times\mathbb{R}^{d} \rightarrow
\mathbb{R}$ such that
for every ${\mathbf x} \in\mathbb{R}^{d}$, $f(\cdot,{\mathbf x})$ is a convex
function. Then, for any
random vector ${\mathbf X} \in\mathbb{R}^{d}$, a subgradient of the map
${\mathbf x} \mapsto E\{f({\mathbf
x},{\mathbf X})\}$ is called the $M$-distribution function of ${\mathbf X}$
with respect to $f$. If
$f({\mathbf x},{\mathbf y}) = \|{\mathbf x} - {\mathbf y}\| - \|{\mathbf y}\|$, where $\|
\cdot\|$ is the usual Euclidean norm, the \mbox{$M$-}distribution function is
the spatial distribution function, whose value at ${\mathbf x}$ with
respect to the probability distribution of ${\mathbf X}$ is $E\{({\mathbf x} -
{\mathbf X})/\|{\mathbf x} - {\mathbf X}\|\}$. If $d=1$, the spatial distribution
simplifies to $2F(x) - 1$, where $F$ is the cumulative distribution
function of $X$. The author of \cite{Kolt97} showed that under certain
conditions, the $M$-distribution function characterizes the probability
distribution of a random vector like the cumulative distribution
function. In that paper, Glivenko--Cantelli and Donsker-type results
were also proved for the empirical $M$-distribution process. These
results are similar to those obtained for the empirical process
associated with the cumulative distribution function in the finite
dimensional multivariate setting. For probability distributions in the
space of real-valued functions on an interval, a notion of
distribution functional was studied in \cite{BHHN09}. But the authors
of \cite{BHHN09} did not study any Glivenko--Cantelli or Donsker-type
result for the empirical processes associated with their distribution
functionals. Further, there is no natural extension of the cumulative
distribution function for probability distributions in general infinite
dimensional Banach spaces.

In this section, we study the spatial distribution in infinite
dimensional Banach
spaces and obtain Glivenko--Cantelli and Donsker-type results for the
associated empirical
processes. Let ${\mathcal X}$ be a smooth Banach space, that is, the
norm function in ${\mathcal X}$ is
G\^ateaux differentiable at each nonzero ${\mathbf x} \in{\mathcal X}$
with derivative, say,
$\SGN_{{\mathbf x}} \in{\mathcal X}^{*}$ (see, e.g., \cite{BV10}). Here,
${\mathcal X}^{*}$ is the dual
space of ${\mathcal X}$, that is, the Banach space of all continuous
real-valued linear functions on
${\mathcal X}$. Thus, $\SGN_{{\mathbf x}}({\mathbf h}) = \lim_{t \rightarrow0}
t^{-1}(\|{\mathbf x} + t{\mathbf
h}\| - \|{\mathbf x}\|)$ for ${\mathbf h} \in{\mathcal X}$. If this limit is
uniform over the set $\{{\mathbf
h} \in{\mathcal X}\dvtx  \|{\mathbf h}\| = 1\}$, then the norm is said to be
Fr\'echet differentiable. If
${\mathcal X}$ is a Hilbert space, $\SGN_{{\mathbf x}} = {\mathbf x}/\|{\mathbf x}\|
$. When ${\mathcal X} =
L_{p}[a,b]$ for some $p \in(1,\infty)$, which is the Banach space of
all functions ${\mathbf x}\dvtx
[a,b] \rightarrow\mathbb{R}$ satisfying $\int_{a}^{b} |{\mathbf
x}(s)|^{p}\,ds < \infty$, then
$\SGN_{{\mathbf x}}({\mathbf h}) = \int_{a}^{b} \sign\{{\mathbf x}(s)\}|{\mathbf
x}(s)|^{p-1}{\mathbf h}(s)\,ds/\|{\mathbf
x}\|^{p-1}$ for all ${\mathbf h} \in{\mathcal X}$. The norm in any Hilbert
space as well as in
$L_{p}[a,b]$ for a $p \in(1,\infty)$ is actually Fr\'echet
differentiable. As a convention, we
define $\SGN_{{\mathbf x}} = {\mathbf0}$ if ${\mathbf x} = {\mathbf0}$.

Let ${\mathbf X}$ be a random element in ${\mathcal X}$, and denote
its probability
distribution by $\mu$. The spatial distribution at ${\mathbf x} \in
{\mathcal X}$ with respect to $\mu$
is defined as $S_{{\mathbf x}} = E\{\SGN_{{\mathbf x} - {\mathbf X}}\}$. Throughout
this article, the
expectations of Banach valued random variables will be defined in the
Bochner sense (see, e.g.,
\cite{AG80}, page~100). The empirical spatial distribution can be defined
as $\widehat{S}_{{\mathbf
x}} = n^{-1} \sum_{i=1}^{n} \SGN_{{\mathbf x} - {\mathbf X}_{i}}$, where ${\mathbf
X}_{1}, {\mathbf X}_{2},\ldots, {\mathbf X}_{n}$ are i.i.d. observations from a probability
distribution $\mu$ in ${\mathcal
X}$. The empirical spatial distribution has been used to develop
Wilcoxon--Mann--Whitney-type
tests for two sample problems in infinite dimensional spaces (see \cite
{CC14c}).

Associated with the spatial distribution is the corresponding
empirical spatial distribution process $\{\widehat{S}_{{\mathbf x}} -
S_{{\mathbf x}}\dvtx  {\mathbf x} \in I\}$ indexed by $I \subseteq{\mathcal X}$.
This is a Banach space valued stochastic process indexed by the
elements in a Banach space. When ${\mathcal X} = \mathbb{R}^{d}$
equipped with the Euclidean norm, the Glivenko--Cantelli and the
Donsker-type results hold for the empirical spatial distribution
process with $I = \mathbb{R}^{d}$ (see~\cite{Kolt97}). The following
theorem states a Glivenko--Cantelli and a Donsker-type result for the
empirical spatial distribution process in a separable Hilbert space.
%
\begin{teo} \label{teo}
Let ${\mathcal X}$ be a separable Hilbert space, and ${\mathcal Z}$ be
a finite dimensional subspace of ${\mathcal X}$. Then $\widehat
{S}_{{\mathbf x}}$ converges to $S_{{\mathbf x}}$ uniformly in ${\mathcal Z}$
in the weak topology of ${\mathcal X}$ almost surely. Further,\vspace*{2pt} if $\mu$
is nonatomic, then for any $d \geq1$ and any continuous linear map
${\mathbf g}\dvtx  {\mathcal X} \rightarrow\mathbb{R}^{d}$, the process $\{
{\mathbf g}(\sqrt{n}(\widehat{S}_{{\mathbf x}} - S_{{\mathbf x}}))\dvtx  {\mathbf x} \in
{\mathcal Z}\}$ converges weakly to a $d$-variate Gaussian process on
${\mathcal Z}$.
\end{teo}
The Glivenko--Cantelli and the Donsker-type results in \cite{Kolt97}
for the empirical spatial distribution process in $\mathbb{R}^{d}$
follow from the above theorem as a straightforward corollary. The
result stated in Theorem~\ref{teo} is also true in Banach spaces like
$L_{p}$ spaces for some even integer $p > 2$ (see the remark after the
proof of Theorem~\ref{teo} given in the \hyperref[sec5]{Appendix}).

A probability measure in an infinite dimensional separable
Banach space ${\mathcal X}$ (e.g., a nondegenerate Gaussian measure)
may assign zero probability to all finite dimensional subspaces.
However, since ${\mathcal X}$ is separable, for any $\varepsilon> 0$,
we can find a compact set $K \subseteq{\mathcal X}$ such that $\mu(K)
> 1 - \varepsilon$ (see, e.g., \cite{AG80}). Thus, given any measurable
set $V \subseteq{\mathcal X}$, there exists a compact set such that
the probability content of $V$ outside this compact set is as small as
we want. The next theorem gives the asymptotic properties of the
empirical spatial distribution process uniformly over any compact
subset of ${\mathcal X}$. We state an assumption that is required for
the next theorem.

{\renewcommand{\theass}{(A)}
\begin{ass}\label{assA}
There exists a map $T({\mathbf x})\dvtx   {\mathcal X}\setminus\{{\mathbf0}\} \rightarrow(0,\infty)$, which is
measurable with respect to the usual Borel $\sigma$-field of ${\mathcal
X}$, and for all ${\mathbf x} \neq{\mathbf0}, -{\mathbf h}$, we have $\|\SGN_{{\mathbf
x} + {\mathbf h}} - \SGN_{{\mathbf x}}\|  \leq T({\mathbf x})\|{\mathbf h}\|$.
\end{ass}}%

Assumption \ref{assA} holds if ${\mathcal X}$ is a Hilbert space or
a $L_{p}$ space for some $p \in[2,\infty)$, and in the former case we
can choose $T({\mathbf x}) = 2/\|{\mathbf x}\|$. For any set $A \subset
{\mathcal X}$, we denote by $N(\varepsilon,A)$ the minimum number of
open balls of radii $\varepsilon$ and centers in~$A$ that are needed to
cover $A$.
%
\begin{teo} \label{teo2}
Let ${\mathcal X}^{*}$ be a separable Banach space, and $K \subseteq
{\mathcal X}$ be a compact set.
\begin{longlist}[(a)]
\item[(a)] Suppose that Assumption \ref{assA} holds, and $\sup_{\|{\mathbf x}\| \leq C}
E_{\mu_{1}} \{T({\mathbf x} - {\mathbf X})\} < \infty$ for each $C > 0$, where
$\mu_{1}$ is the nonatomic part of $\mu$. Then $\widehat{S}_{{\mathbf x}}$
converges to $S_{{\mathbf x}}$ uniformly over ${\mathbf x} \in K$ in the norm
topology of ${\mathcal X}^{*}$ almost surely.\vadjust{\goodbreak}

\item[(b)] Let $\mu$ be a nonatomic probability measure, Assumption \ref{assA} hold,
and $\sup_{\|{\mathbf x}\| \leq C} E_{\mu} \{T^{2}({\mathbf x} - {\mathbf X})\} <
\infty$ for each $C > 0$. If $\int_{0}^{1} {\sqrt{\ln  N(\varepsilon,K)}\,d{\varepsilon}} < \infty$ for each $\varepsilon> 0$, then the
empirical process $\widehat{{\mathbf S}}_{{\mathbf g}} = \{{\mathbf g}(\sqrt
{n}(\widehat{S}_{{\mathbf x}} - S_{{\mathbf x}}))\dvtx  {\mathbf x} \in K\}$ converges
weakly to a $d$-variate Gaussian process on $K$ for any $d \geq1$ and
any continuous linear function ${\mathbf g}\dvtx  {\mathcal X}^{*} \rightarrow
\mathbb{R}^{d}$. Further, if ${\mathcal X}$ is a separable Hilbert
space, then for any Lipschitz continuous function ${\mathbf g}\dvtx  {\mathcal
X} \rightarrow\mathbb{R}^{d}$, $\widehat{{\mathbf S}}_{{\mathbf g}}$ converges
weakly to a $\mathbb{R}^d$-valued stochastic process on $K$.
\end{longlist}
\end{teo}

If $\mu$ is a purely atomic measure, the Glivenko--Cantelli-type result
in part~(a) of the above theorem holds over the entire space ${\mathcal
X}$ (see Lemma~\ref{lemma0} in the \hyperref[sec5]{Appendix}). It follows from part~(a)
of the above theorem and the tightness of any probability measure in
any complete separable metric space that $\int_{{\mathcal X}} \|\widehat
{S}_{{\mathbf x}} - S_{{\mathbf x}}\|^{2} \mu(d{\mathbf x}) \rightarrow0$ as $n
\rightarrow\infty$ \emph{almost surely}. If we choose $d=1$ and ${\mathbf
g}({\mathbf x}) = \|{\mathbf x}\|$ in the second statement in part~(b) of the\vspace*{2pt}
above theorem, it follows that $\sup_{{\mathbf x} \in K} \|\widehat
{S}_{{\mathbf x}} - S_{{\mathbf x}}\| = O_{P}(1/\sqrt{n})$ and $\int_{{\mathcal
X}} \|\widehat{S}_{{\mathbf x}} - S_{{\mathbf x}}\|^{2} \mu(d{\mathbf x}) =
O_{P}(1/n)$ as $n \rightarrow\infty$.

Let ${\mathcal X}$ be a separable Hilbert space and ${\mathbf X} =
\sum_{k=1}^{\infty} X_{k}\psi_{k}$ for an orthonormal basis $\{\psi
_{k}\}_{k \geq1}$ of ${\mathcal X}$. Then the moment condition assumed
in part~(a) [resp., part~(b)] of the above theorem holds if some
bivariate (resp., trivariate) marginal of $(X_{1}, X_{2},\ldots)$ has a
density under $\mu_{1}$ (resp., $\mu$) that is bounded on bounded
subsets of $\mathbb{R}^{2}$ (resp., $\mathbb{R}^{3}$).

Let $J = \int_{0}^{1}{\sqrt{\ln N(\varepsilon,K)}\,d\varepsilon}$.
It is easy to verify that $J < \infty$ for every compact set $K$ in any
finite dimensional Banach space. The finiteness of $J$ is also true for
many compact sets in various infinite dimensional function spaces like
the compact sets in $L_{p}$ spaces for $p \in(1,\infty)$ whose
elements have continuous partial derivatives up to order $r-1$ for some
$r \geq1$ and the $r$th order partial derivatives are Holder
continuous with a positive exponent (see, e.g.,~\cite{KT61}).

\section{Spatial quantiles in Banach spaces}\label{sec4}
 An important property of the spatial distribution in finite
dimensional Euclidean spaces is its strict monotonicity for a class of
nonatomic probability distributions. This along with its continuity and
the surjective property was used to define the spatial quantile as the
inverse of the spatial distribution in these spaces (see \cite
{Kolt97}). The following result shows that even in a class of infinite
dimensional Banach spaces, we have the strict monotonicity, the
surjectivity and the continuity of the spatial distribution map. A
Banach space ${\mathcal X}$ is said to be strictly convex if for any
${\mathbf x} \neq{\mathbf y} \in{\mathcal X}$ satisfying $\|{\mathbf x}\| = \|
{\mathbf y}\| = 1$, we have $\|({\mathbf x} + {\mathbf y})/2\| < 1$ (see, e.g.,~\cite{BV10}). Hilbert spaces and $L_{p}$ spaces for $p \in(1,\infty)$
are strictly convex. A line in ${\mathcal X}$ through ${\mathbf x}_{1}$ and
${\mathbf x}_{2}$ is defined as the set of points $\{a{\mathbf x}_{1} +
(1-a){\mathbf x}_{2}\dvtx  a \in\mathbb{R}\}$.
%
\begin{teo} \label{teo1}
Let ${\mathcal X}$ be a smooth, strictly convex Banach space, and
suppose that $\mu$ is nonatomic probability measure in ${\mathcal X}$.
If $\mu$ is not entirely supported on a line in ${\mathcal X}$, the
spatial distribution map ${\mathbf x} \mapsto S_{{\mathbf x}}$ is strictly
monotone, that is, $(S_{{\mathbf x}} - S_{{\mathbf y}})({\mathbf x} - {\mathbf y}) > 0$
for all ${\mathbf x}, {\mathbf y} \in{\mathcal X}$ with ${\mathbf x} \neq{\mathbf
y}$. The range of the spatial distribution map is the entire open unit
ball in ${\mathcal X}^{*}$ if ${\mathcal X}$ is reflexive (i.e.,
${\mathcal X} = {\mathcal X}^{**}$). If the norm in ${\mathcal X}$ is
Fr\'echet differentiable, the spatial distribution map is continuous.
\end{teo}
Under the conditions of Theorem~\ref{teo1}, for any ${\mathbf u}$ in the
open unit ball ${\mathcal B}^{*}({\mathbf0},1)$ in ${\mathcal X}^{*}$, the
spatial ${\mathbf u}$-quantile ${\mathbf Q}({\mathbf u})$ can be defined as the
inverse, evaluated at ${\mathbf u}$, of the spatial distribution map. Thus,
${\mathbf Q}({\mathbf u})$ is the solution of the equation $E\{\SGN_{{\mathbf Q} -
{\mathbf X}}\} = {\mathbf u}$. When $\mu$ has atoms, we can define ${\mathbf
Q}({\mathbf u})$ by \emph{appropriately inverting} the spatial distribution
map, which is now a continuous bijection from ${\mathcal X}{\setminus
}A_{\mu}$ to ${\mathcal B}^{*}({\mathbf0},1){\setminus}\bigcup_{{\mathbf x}
\in A_{\mu}} \overline{{\mathcal B}}^{*}(S_{{\mathbf x}},p({\mathbf x}))$ if
the other conditions in Theorem~\ref{teo1} hold but it is discontinuous
at each ${\mathbf x} \in A_{\mu}$. Here, $A_{\mu}$ denotes the set of atoms
of~$\mu$, $p({\mathbf x}) = P({\mathbf X} = {\mathbf x})$ for ${\mathbf x} \in A_{\mu
}$, and ${\mathcal B}^{*}({\mathbf z},r)$ and $\overline{{\mathcal
B}}^{*}({\mathbf z},r)$ denote the open and the closed balls in ${\mathcal X}^{*}$,
respectively, with radius $r$ and center ${\mathbf z} \in{\mathcal
X}^{*}$. Even if $\mu$ has atoms,
it can be shown that ${\mathbf Q}({\mathbf u})$ is the minimizer of $E\{\|{\mathbf
Q} - {\mathbf X}\| - \|{\mathbf
X}\|\} - {\mathbf u}({\mathbf Q})$ with respect to ${\mathbf Q} \in{\mathcal X}$.
Spatial quantiles have been
defined in $\mathbb{R}^{d}$ through such a minimization problem in \cite
{Chau96} and
\cite{Kolt97}. The former paper also mentioned about the extension of
spatial quantiles into
general Banach spaces. The properties of spatial quantiles for
probability distributions in
$\mathbb{R}^{d}$ equipped with the $l_{p}$-norm for some $p \in
[1,\infty)$ was studied by
\cite{Chak01}. Suppose that we have a unimodal probability density
function in $\mathbb{R}^{d}$. If the density function is a strictly
decreasing function of $\|{\mathbf x}\|_{p}$, where $\|\cdot\|_{p}$ is the
$l_{p}$-norm, then it can be easily shown that the density contours and
the contours of the spatial quantiles computed using the $l_{p}$-norm
coincide.

Note that the central quantiles correspond to small values of
$\|{\mathbf u}\|$, while the
extreme quantiles correspond to larger values of $\|{\mathbf u}\|$.
Further, ${\mathbf u}/\|{\mathbf u}\|$
gives the direction of the proximity/remoteness of ${\mathbf Q}({\mathbf u})$
relative the center of
the probability distribution. For example, let ${\mathbf X} =
(X_{1},X_{2},\ldots)$ be a
nondegenerate \mbox{random} element symmetric about zero in $l_{p}$ for some
$p \in(1,\infty)$. So, the spatial median of ${\mathbf X}$ is zero. For
any ${\mathbf u}$ in the open unit ball of $l_{q} = l_{p}^{*}$, where $1/p
+ 1/q = 1$, the spatial ${\mathbf u}$-quantile ${\mathbf Q}({\mathbf u}) =
(q_{1},q_{2},\ldots)$ of ${\mathbf X}$ will satisfy the equation $E\{
\sign(q_{k} - X_{k})|q_{k} -
X_{k}|^{p-1}/\|{\mathbf Q}({\mathbf u}) - {\mathbf X}\|^{p-1}\} = u_{k}$ for all $k
\geq1$. If $\|{\mathbf
u}\|$ is close to zero, then it follows from the symmetry of the
distribution of $X_{k}$
that $q_{k}$ should also be close to zero for all $k \geq1$. Further,
if the $q_{k}$'s
are large, the corresponding ${\mathbf Q}({\mathbf u})$ is an extreme quantile
of the distribution of
${\mathbf X}$.

 The spatial quantile possesses an equivariance property under
the class of affine
transformations $L\dvtx  {\mathcal X} \rightarrow{\mathcal X}$ of the form
$L({\mathbf x}) = cA({\mathbf x}) +
{\mathbf a}$, where $c > 0$, ${\mathbf a} \in{\mathcal X}$ and $A\dvtx  {\mathcal
X} \rightarrow{\mathcal X}$ is a
linear surjective isometry, that is, $\|A({\mathbf x})\| = \|{\mathbf x}\|$ for
all ${\mathbf x} \in{\mathcal
X}$. Using the surjective property of $A$ it follows that minimizing
$E\{\|{\mathbf Q} - L({\mathbf
X})\| - \|L({\mathbf X})\|\} - {\mathbf u}({\mathbf Q})$ over ${\mathbf Q} \in
{\mathcal X}$ is equivalent to
minimizing $E\{\|A({\mathbf Q}') - A({\mathbf X})\| - \|A({\mathbf X})\|\} - {\mathbf
u}(A({\mathbf Q}'))$ over
${\mathbf Q}' \in{\mathcal X}$, where ${\mathbf Q} = L({\mathbf Q}')$. The last
minimization problem is the
same as minimizing $E\{\|{\mathbf Q}' - {\mathbf X}\| - \|{\mathbf X}\|\} -
(A^{*}({\mathbf u}))({\mathbf Q}')$
over ${\mathbf Q}' \in{\mathcal X}$ by virtue of the isometry of $A$.
Here, $A^{*}\dvtx  {\mathcal X}^{*}
\rightarrow{\mathcal X}^{*}$ denotes the adjoint of $A$ (see, e.g.,
\cite{FHH01}). Thus, the
spatial ${\mathbf u}$-quantile of the distribution of $L({\mathbf X})$ equals
$L({\mathbf Q}(A^{*}({\mathbf
u})))$, where ${\mathbf Q}(A^{*}({\mathbf u}))$ is the $A^{*}({\mathbf
u})$-quantile of the distribution of
${\mathbf X}$.

 The sample spatial ${\mathbf u}$-quantile can be defined as the
minimizer over ${\mathbf Q} \in{\mathcal X}$ of $n^{-1} \sum_{i=1}^{n} \{\|
{\mathbf Q} - {\mathbf X}_{i}\| - \|{\mathbf X}_{i}\|\} - {\mathbf u}({\mathbf Q})$. Note
that this minimization problem is an infinite dimensional one and is
intractable in general. The author of \cite{Cadr01} proposed an
alternative estimator of the spatial median (i.e., when ${\mathbf u} = {\mathbf
0}$) by considering the above empirical minimization problem only over
the data points. However, as mentioned by that author, this estimator
will be inconsistent when the population spatial median lies outside
the support of the distribution. The author of \cite{Gerv08} proposed
an algorithm for computing the sample spatial median in Hilbert spaces.
However, the idea does not extend to spatial quantiles or into general
Banach spaces.

 We shall now discuss a computational procedure for sample
spatial quantiles in a Banach
space. We assume that ${\mathcal X}$ is a Banach space having a
Schauder basis
$\{\phi_{1},\phi_{2},\ldots\}$, say, so that for any ${\mathbf x} \in
{\mathcal X}$, there exists a
unique sequence of real numbers $\{x_{k}\}_{k \geq1}$ such that ${\mathbf
x} = \sum_{k=1}^{\infty}
x_{k}\phi_{k}$ (see, e.g., \cite{FHH01}). Note that if ${\mathcal X}$
is a Hilbert space and
$\{\phi_{1},\phi_{2},\ldots\}$ is an orthonormal basis of ${\mathcal
X}$, then it is a Schauder
basis of~${\mathcal X}$. Let ${\mathcal Z}_{n} = \span\{\phi_{1},\phi
_{2}, \ldots,\phi_{d(n)}\}$, where
$d(n)$ is a positive integer depending on the sample size $n$. Define
${\mathbf z}^{(n)} =
\sum_{k=1}^{d(n)} a_{k}\phi_{k}$, where ${\mathbf z} = \sum_{k=1}^{\infty}
a_{k}\phi_{k}$. We will
assume that $\|{\mathbf z}^{(n)}\| \leq\|{\mathbf z}\|$ for all $n \geq1$ and
${\mathbf z} \in{\mathcal X}$.
Note that if ${\mathcal X}$ is a Hilbert space, and $\{\phi_{1},\phi
_{2},\ldots\}$ is an
orthonormal basis of ${\mathcal X}$, then ${\mathbf z}^{(n)}$ is the
orthogonal projection of ${\mathbf z}$
onto ${\mathcal Z}_{n}$. For each $k \geq1$, define $\widetilde{\phi
}_{k}$ to be the continuous
linear functional on ${\mathcal X}$ given by $\widetilde{\phi}_{k}({\mathbf
z}) = a_{k}$. Let us assume
that $\{\widetilde{\phi}_{1},\widetilde{\phi}_{2},\ldots\}$ is a
Schauder basis of ${\mathcal
X}^{*}$. Define ${\mathbf u}^{(n)} = \sum_{k=1}^{d(n)} b_{k}\widetilde{\phi
}_{k}$, where ${\mathbf u}
\in{\mathcal B}^{*}({\mathbf0},1)$ and ${\mathbf u} = \sum_{k=1}^{\infty}
b_{k}\widetilde{\phi}_{k}$. We
also assume that $\|{\mathbf u}^{(n)}\| \leq\|{\mathbf u}\|$ for all $n \geq
1$ and ${\mathbf u} \in{\mathcal
B}^{*}({\mathbf0},1)$. The above assumptions concerning the Schauder bases
of a Banach space and
its dual space hold for any separable Hilbert space and any $L_{p}$
space with $p \in
(1,\infty)$ (see,\vspace*{1pt} e.g., \cite{FHH01}, pages~166--169). We compute the
sample spatial ${\mathbf
u}$-quantile $\widehat{{\mathbf Q}}({\mathbf u})$ as the minimizer of
$n^{-1}\sum_{i=1}^{n} \{\|{\mathbf Q}
- {\mathbf X}_{i}^{(n)}\| - \|{\mathbf X}_{i}^{(n)}\|\} - {\mathbf u}^{(n)}({\mathbf
Q})$ over ${\mathbf Q} \in
{\mathcal Z}_{n}$.

 For all the numerical studies in our paper, we have chosen
$d(n) = [\sqrt{n}]$. In our
simulated data examples, sample quantiles computed with this choice of
$d(n)$ approximate the
true quantiles quite well. We will later show that this choice of
$d(n)$ ensures the
consistency of sample quantiles in a class of Banach spaces, and is
sufficient to prove their asymptotic Gaussianity in separable Hilbert
spaces (cf. Theorems~\ref{teo5} and~\ref{teo7}).

\begin{figure}

\includegraphics{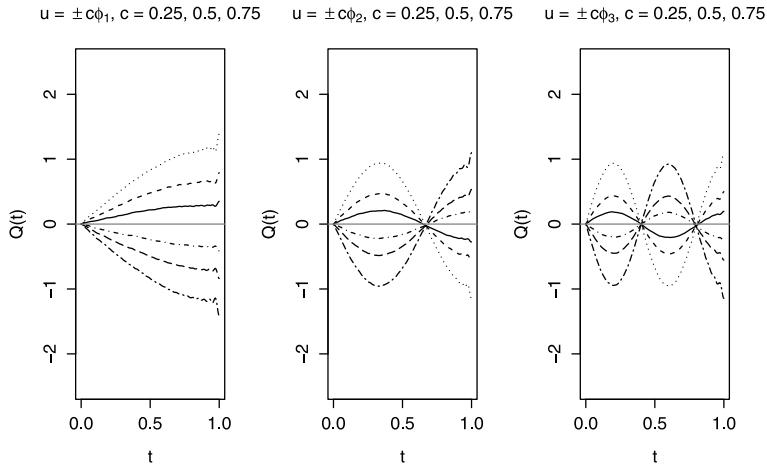}

\caption{The plots of the spatial quantiles of the standard Brownian
motion including the spatial median (horizontal line through zero in
all the plots). For each $k = 1,2,3$, the spatial quantiles
corresponding to ${\mathbf u} = c\phi_{k}$ for $c = 0.25$, $0.5$ and $0.75$
are given by the solid (---), the dashed (- - -) and the dotted ($\cdots$) curves, respectively, while those corresponding to ${\mathbf u} = -c\phi
_{k}$ for these $c$ values are given by the dot-dashed (-- $\cdot$ --),
the long-dashed (-- --) and the two-dashed (-- - --) curves, respectively.}\label{Fig1}
\end{figure}

 We now demonstrate the spatial quantiles using some simulated
and real data. We have
considered the random element ${\mathbf X} = \sum_{k=1}^{\infty} \lambda
_{k}Y_{k}\phi_{k}$ in
$L_{2}[0,1]$. Here, the $Y_{k}$'s are independent $N(0,1)$ random
variables, $\lambda_{k} =
\{(k-0.5)\pi\}^{-1}$ and $\phi_{k}(t) = \sqrt{2}\sin\{(k-0.5){\pi}t\}$
for $k \geq1$. Note that
${\mathbf X}$ has the distribution of the standard Brownian motion on
$[0,1]$ with $\phi_{k}$ being
the eigenfunction associated with the eigenvalue $\lambda_{k}^{2}$ of
the covariance kernel of
the standard Brownian motion. We have first plotted the population
spatial quantiles of the
standard Brownian motion for ${\mathbf u} = \pm c\phi_{k}$, where $k =
1,2,3$ and $c = 0.25, 0.5,
0.75$ (see Figure~\ref{Fig1}). Note that $\lambda_{1}Y_{1}$, $\lambda
_{2}Y_{2}$ and
$\lambda_{3}Y_{3}$ account for $81.1\%$, $9\%$ and $3.24\%$,
respectively, of the total
variation $E(\|{\mathbf X}\|^{2}) = \sum_{k=1}^{\infty} \Var(\lambda
_{k}Y_{k}) = \sum_{k=1}^{\infty}
\lambda_{k}^{2}$ in the Brownian motion process. For computing the
population spatial
quantiles, we generated a large sample of size $n = 2500$ from the
standard Brownian motion and
computed the sample spatial quantiles with $d(n) = [\sqrt{n}]$ and
${\mathcal Z}_{n} =
\span\{\phi_{1},\phi_{2},\ldots,\phi_{d(n)}\}$.

%

%
\begin{figure}

\includegraphics{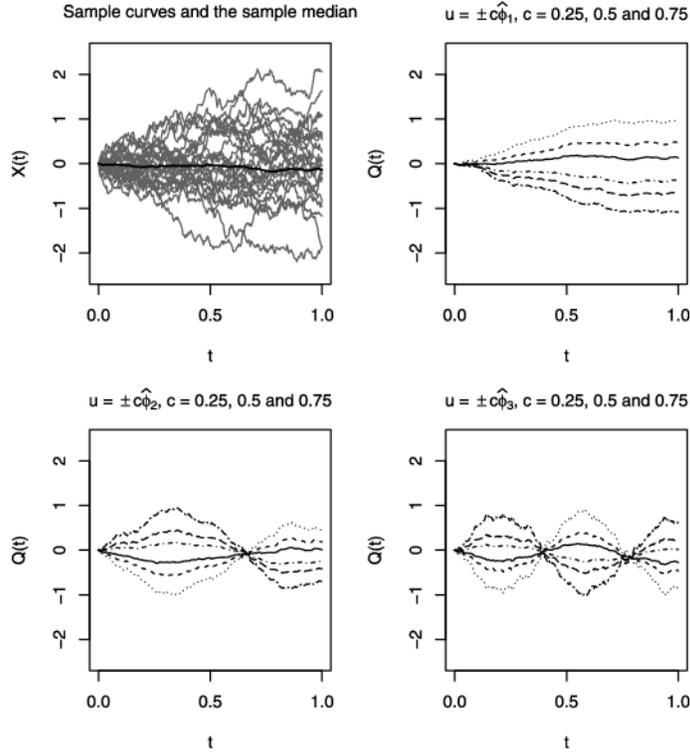}

\caption{The plots of the simulated data along with the sample spatial
median (bold curve in the top left plot), and other sample spatial
quantiles (in the remaining plots). For each $k = 1,2,3$, the sample
spatial quantiles corresponding to ${\mathbf u} = c\phi_{k}$ for $c =
0.25$, $0.5$ and $0.75$ are given by the solid (---), the dashed (- -
-) and the dotted ($\cdots$) curves, respectively, while those
corresponding to ${\mathbf u} = -c\phi_{k}$ for these $c$ values are given
by the dot-dashed (-- $\cdot$ --), the long-dashed (-- --) and the
two-dashed (-- - --) curves, respectively.}
\label{Fig2}
\end{figure}

\begin{figure}

\includegraphics{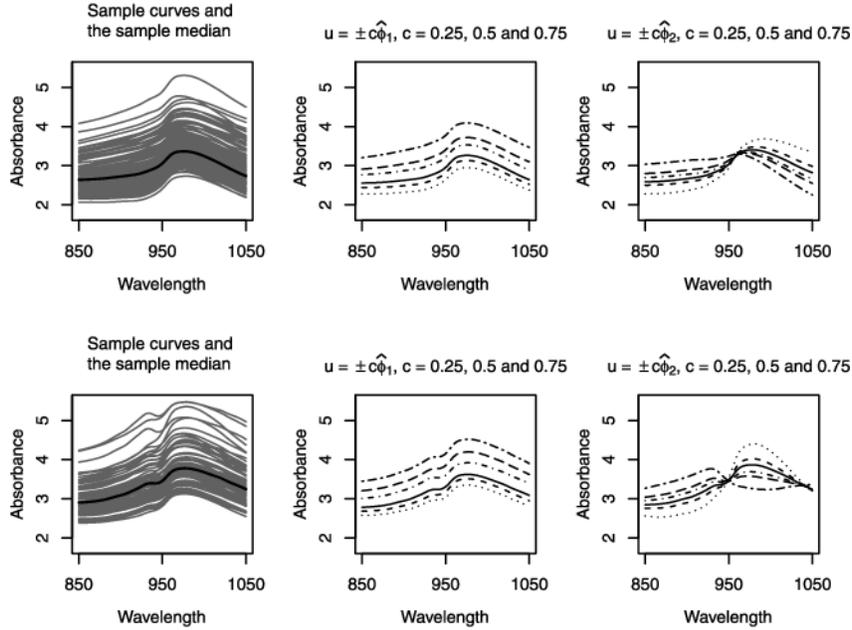}

\caption{The plots of the spectrometric data and the sample spatial
quantiles. The plots in the first column show the observations for fat
content $\leq$20\% and $>$20\% along with the sample spatial medians
(bold curves). For each $k = 1,2$, the sample spatial quantiles
corresponding to ${\mathbf u} = c\phi_{k}$ for $c = 0.25$, $0.5$ and $0.75$
are given by the solid (---), the dashed (- - -) and the dotted ($\cdots
$) curves, respectively, while those corresponding to ${\mathbf u} = -c\phi
_{k}$ for these $c$ values are given by the dot-dashed (-- $\cdot$ --),
the long-dashed (-- --) and the two-dashed (-- - --) curves,
respectively, in the plots in the second and the third columns.}
\label{Fig3}
\end{figure}

 Our simulated data consists of $n=50$ sample curves from the
standard Brownian motion,
and each sample curve is observed at $250$ equispaced points in
$[0,1]$. The real dataset
considered here is available at \url{http://www.math.univ-toulouse.fr/\textasciitilde staph/npfda/},
and it contains the spectrometric curves of $n = 215$ meat units
measured at $100$ wavelengths
in the range $850$ nm to $1050$ nm along with the fat\vadjust{\goodbreak} content of each
unit categorized into two
classes, namely, below and above $20\%$. The sample curves of the real
data may be viewed as
elements in $L_{2}[850,1050]$ equipped with its usual norm. For each of
the simulated and the
real dataset, we have chosen $d(n) = [\sqrt{n}]$, and ${\mathcal
Z}_{n}$ is constructed using the
eigenvectors associated with the $d(n)$ largest eigenvalues of the
sample covariance matrix.
For computing the sample spatial quantiles for both the simulated and
the real data, we have
first computed the sample spatial quantiles for the centered data
obtained by subtracting the
sample mean from each observation, and then added back the sample mean
to~the computed sample
spatial quantiles. Figure~\ref{Fig2} (resp., Figure~\ref{Fig3}) shows
the plots of the
simulated dataset (resp., real dataset) along with the sample spatial
median and the
sample spatial quantiles corresponding to ${\mathbf u} = \pm c\widehat{\phi
}_{k}$ for $k = 1,2,3$
($k = 1,2$), where $c = 0.25, 0.5, 0.75$ and $\widehat{\phi}_{k}$ is
the eigenvector associated
with the $k$th largest eigenvalue of the sample covariance matrix for
$k \geq1$. The
percentage of the total variation in the simulated data explained by
the first three sample
eigenvectors is almost same as the population values mentioned earlier.
For each of the two
classes in the real dataset, the first two sample eigenvectors account
for about $99.5\%$ of
the total variation in that class.

%

 For each $k$, the spatial ${\mathbf u}$-quantiles of the standard
Brownian motion corresponding to ${\mathbf u} = c\phi_{k}$ and $-c\phi_{k}$
exhibit an ordering, where the spatial ${\mathbf u}$-quantile associated
with a smaller $c$ value is relatively closer to the spatial median
than the spatial ${\mathbf u}$-quantile associated with a larger $c$ value
(see Figure~\ref{Fig1}). A similar ordering is also seen for the
sample spatial quantiles of both the simulated and the two classes in
the real dataset. The sample spatial median for the simulated data is
close to the zero function (see Figure~\ref{Fig2}), which is the
spatial median of the standard Brownian motion. There is a noticeable
difference in the locations of the sample spatial median and the sample
spatial quantiles corresponding to ${\mathbf u} = \pm c\widehat{\phi}_{1}$
between the two classes in the real dataset (see Figure~\ref{Fig3}).
Moreover, the sample spatial quantiles of the two classes in the real
dataset are different in their shapes.

\subsection{Asymptotic properties of sample spatial quantiles}\label{subsec42}

The\vspace*{2pt} following theorem gives the strong consistency of $\widehat
{{\mathbf Q}}({\mathbf u})$ in the norm topology for a class of Banach spaces.
The norm in a Banach space ${\mathcal X}$ is said to be locally
uniformly rotund if for any sequence $\{{\mathbf x}_{n}\}_{n \geq1} \in
{\mathcal X}$ and any ${\mathbf x} \in{\mathcal X}$ satisfying $\|{\mathbf
x}_{n}\| = \|{\mathbf x}\| = 1$ for all $n \geq1$, $\lim_{n \rightarrow
\infty} \|{\mathbf x}_{n} + {\mathbf x}\| = 2$ implies $\lim_{n \rightarrow
\infty} \|{\mathbf x}_{n} - {\mathbf x}\| = 0$ (see, e.g., \cite{BV10}). The
norm in any Hilbert space or any $L_{p}$ space for $p \in(1,\infty)$
is locally uniformly rotund.
%
\begin{teo} \label{teo5}
Suppose that ${\mathcal X}$ is a separable, reflexive Banach space such
that the norm in ${\mathcal X}$ is locally uniformly rotund, and assume
that $\mu$ is nonatomic and not entirely supported on a line in
${\mathcal X}$. Then $\|\widehat{{\mathbf Q}}({\mathbf u}) - {\mathbf Q}({\mathbf u})\|
\rightarrow0$ as $n \rightarrow\infty$ \emph{almost surely} if $d(n)
\rightarrow\infty$ as $n \rightarrow\infty$.
\end{teo}

Since $\widehat{{\mathbf Q}}({\mathbf u})$ is a nonlinear function of
the data, in order to study its asymptotic distribution, we need to
approximate it by a suitable linear function of the data. In finite
dimensions, this is achieved through a Bahadur-type asymptotic linear
representation (see, e.g., \cite{Chau96} and \cite{Kolt97}), and our
next theorem gives a similar representation in infinite dimensional
Hilbert spaces. Consider the real-valued function $g({\mathbf Q}) = E\{\|
{\mathbf Q} - {\mathbf X}\| - \|{\mathbf X}\|\} - {\mathbf u}({\mathbf Q})$ defined on a
Hilbert space ${\mathcal X}$, and denote its Hessian at ${\mathbf Q} \in
{\mathcal X}$ by $J_{{\mathbf Q}}$, which is a symmetric bounded bilinear
function from ${\mathcal X} \times{\mathcal X}$ into $\mathbb{R}$ satisfying
\begin{eqnarray*}
&& \lim_{t \rightarrow0} \biggl\llvert g({\mathbf Q} + t{\mathbf h}) - g({\mathbf Q})
- tE \biggl\{\frac{{\mathbf Q}-{\mathbf X}}{\|{\mathbf Q}-{\mathbf X}\|} - {\mathbf u} \biggr\} ({\mathbf h}) -
\frac{t^{2}}{2}J_{{\mathbf Q}}({\mathbf h},{\mathbf h})\biggr\rrvert \Big/t^{2}
= 0
\end{eqnarray*}
for any ${\mathbf h} \in{\mathcal X}$. We define the continuous linear
operator $\widetilde{J}_{{\mathbf Q}}\dvtx  {\mathcal X} \rightarrow{\mathcal
X}$ associated with $J_{{\mathbf Q}}$ by the equation $\langle\widetilde
{J}_{{\mathbf Q}}({\mathbf h}),{\mathbf v}\rangle= J_{{\mathbf Q}}({\mathbf h},{\mathbf v})$
for every ${\mathbf h}, {\mathbf v} \in{\mathcal X}$. We define the Hessian
$J_{n,{\mathbf Q}}$ of the function $g_{n}({\mathbf Q}) = E\{\|{\mathbf Q} - {\mathbf
X}^{(n)}\| - \|{\mathbf X}^{(n)}\|\} - {\mathbf u}^{(n)}({\mathbf Q})$, which is
defined on ${\mathcal Z}_{n}$, in a similar way. The continuous linear
operator associated with $J_{n,{\mathbf Q}}$ is denoted by $\widetilde
{J}_{n,{\mathbf Q}}$. Here, we consider an orthonormal basis of ${\mathcal
X}$ (which is a Schauder basis), and ${\mathcal Z}_{n}$ is as chosen as
in Section~\ref{sec4}. Let ${\mathbf Q}_{n}({\mathbf u}) = \arg\min_{{\mathbf Q}
\in{\mathcal Z}_{n}} g_{n}({\mathbf Q})$ and define $\mbox{B}_{n}({\mathbf u})
= \|{\mathbf Q}_{n}({\mathbf u}) - {\mathbf Q}({\mathbf u})\|$. It can be shown that
$\mbox{B}_{n}({\mathbf u}) \rightarrow0$ as $n \rightarrow\infty$. We
make the
following assumption, which will be required for Theorem~\ref{teo6} below.

{\renewcommand{\theass}{(B)}
\begin{ass}\label{assB}
Suppose that $\mu$ is nonatomic
and not entirely supported on a line in ${\mathcal X}$, and $\sup_{{\mathbf
Q} \in{\mathcal Z}_{n}, \|{\mathbf Q}\| \leq C} E\{\|{\mathbf Q} - {\mathbf
X}^{(n)}\|^{-2}\} < \infty$ for each $C > 0$ and all appropriately
large $n$.
\end{ass}}%

 As discussed after Assumption \ref{assA} in Section~\ref{sec2}, if
${\mathcal X}$ is a Hilbert space, we can choose $T({\mathbf x}) = 2/\|{\mathbf
x}\|$ in that assumption. Thus, Assumption \ref{assB} can be viewed as a
$d(n)$-dimensional analog of the moment condition assumed in part~(b)
of Theorem~\ref{teo2}. Also, it holds under the same situation as
discussed after Theorem~\ref{teo2}.
%
\begin{teo} \label{teo6}
Let ${\mathcal X}$ be a separable Hilbert space, and Assumption \ref{assB}
hold. Then the following Bahadur-type asymptotic linear representation
holds if for some $\alpha\in(0,1/2]$, $d(n)/n^{1-2\alpha}$ tends to a
positive constant as $n \rightarrow\infty$.
\begin{eqnarray*}
\widehat{{\mathbf Q}}({\mathbf u}) - {\mathbf Q}_{n}({\mathbf u}) &=& -
\frac{1}{n} \sum_{i=1}^{n} [
\widetilde{J}_{n,{\mathbf Q}_{n}({\mathbf u})}]^{-1} \biggl( \frac
{{\mathbf Q}_{n}({\mathbf u}) - {\mathbf X}_{i}^{(n)}}{\|{\mathbf Q}_{n}({\mathbf u}) -
{\mathbf X}_{i}^{(n)}\|} - {\mathbf
u}^{(n)} \biggr) + {\mathbf R}_{n},
\end{eqnarray*}
where ${\mathbf R}_{n} = O((\ln n)/n^{2\alpha})$ as $n \rightarrow\infty$
\emph{almost surely}.
\end{teo}
The Bahadur-type representation of the sample spatial ${\mathbf
u}$-quantile in finite \mbox{dimensional}
Euclidean spaces (see, e.g., \cite{Chau96} and \cite{Kolt97}) can be
obtained as a
straightforward corollary of the above theorem by choosing $\alpha=
1/2$. Under the
assumptions\vspace*{2pt} of the preceding theorem, if $\alpha\in(1/4,1/2]$, we
have the asymptotic
\mbox{Gaussianity} of $\sqrt{n}(\widehat{{\mathbf Q}}({\mathbf u}) - {\mathbf Q}_{n}({\mathbf
u}))$ as $n \rightarrow
\infty$.

 The extension of the above Bahadur-type representation into
general Banach spaces is a
challenging task mainly due to two reasons. First, although some
version of Bernstein-type
exponential bounds as in Fact~\ref{fact1} are available in general
Banach spaces (see, e.g.,
Theorem 2.1 in \cite{Yuri76}), those bounds are not \mbox{adequate} for
extending the proof
from Hilbert spaces into general Banach spaces. Next, the lower bound
of $J_{n,{\mathbf Q}}({\mathbf
h},{\mathbf h})/\|{\mathbf h}\|^{2}$ in Fact~\ref{lemma2} is not always true in
general Banach spaces.
For instance, let ${\mathcal X} = l_{4}$ and ${\mathbf X} =
(X_{1},X_{2},\ldots)$ be a zero mean
Gaussian random element in ${\mathcal X}$. Let ${\mathcal Z}_{n} =
\span\{{\mathbf e}_{1},{\mathbf
e}_{2},\ldots,{\mathbf e}_{d(n)}\}$, where ${\mathbf e}_{k} = (I(j=k)\dvtx j\geq
1)$, $k \geq1$, which form
the canonical Schauder basis for $l_{4}$. Let ${\mathbf h}_{n} = {\mathbf
e}_{d(n)} \in{\mathcal Z}_{n}$.
Then, for any ${\mathbf Q} = (q_{1},q_{2},\ldots) \in{\mathcal Z}_{n}$, it\vspace*{1pt}
can be shown that
$J_{n,{\mathbf Q}}({\mathbf h}_{n},{\mathbf h}_{n})/\|{\mathbf h}_{n}\|^{2} \leq
3E[(q_{d(n)} -
X_{d(n)})^{2}/\|{\mathbf Q} - {\mathbf X}^{(n)}\|^3]$. It can also be shown
that the right-hand side of the preceding inequality converges to zero
as $n \rightarrow\infty$ by observing that $|q_{d(n)} - X_{d(n)}|
\rightarrow0$ almost surely and $d(n) \rightarrow\infty$ as $n
\rightarrow\infty$. This clearly implies that the lower bound in Fact
\ref{lemma2} does not hold in this case.

We\vspace*{1pt} shall now discuss some situations when $\mbox{B}_{n}({\mathbf
u}) = \|{\mathbf Q}_{n}({\mathbf
u}) - {\mathbf Q}({\mathbf u})\|$ satisfies $\lim_{n \rightarrow\infty} \sqrt
{n}\mbox{B}_{n}({\mathbf u})
= 0$. This along with the weak convergence of $\sqrt{n}(\widehat{{\mathbf
Q}}({\mathbf u}) - {\mathbf
Q}_{n}({\mathbf u}))$ stated above will give the asymptotic Gaussianity of
$\sqrt{n}(\widehat{{\mathbf
Q}}({\mathbf u}) - {\mathbf Q}({\mathbf u}))$ as $n \rightarrow\infty$. Under the
assumptions of Theorem
\ref{teo6}, it can be shown that for some constants $b_{1}, b_{2} > 0$,
we have
$\mbox{B}_{n}({\mathbf u}) \leq b_{1}r_{n} + b_{2}s_{n}$ for all\vspace*{1pt} large $n$,
where $r_{n} =
E\{\|{\mathbf X} - {\mathbf X}^{(n)}\|/\|{\mathbf Q}({\mathbf u}) - {\mathbf X}\|\}$ and
$s_{n} = \|{\mathbf u} - {\mathbf
u}^{(n)}\|$. Let us take ${\mathcal X} = L_{2}([a,b],\nu)$, which is
the space of all real-valued
functions ${\mathbf x}$ on $[a,b] \subseteq\mathbb{R}$ with $\nu$ a
probability measure on $[a,b]$
such that $\int{\mathbf x}^{2}(t)\nu(dt) < \infty$. Suppose ${\mathbf X}$ has
the Karhunen--Lo\`eve
expansion ${\mathbf X} = {\mathbf m} + \sum_{k=1}^{\infty} \lambda_{k}Y_{k}\phi
_{k}$, where the
$Y_{k}$'s are uncorrelated random variables with zero means and unit
variances, the
$\lambda_{k}^{2}$'s and the $\phi_{k}$'s are the eigenvalues and the
eigenfunctions,
respectively, of the covariance of ${\mathbf X}$. Let ${\mathcal Z}_{n} =
\span\{\phi_{1},\phi_{2},\ldots,\phi_{d(n)}\}$. Under the assumptions of
Theorem~\ref{teo6}, it
can be shown that\vadjust{\goodbreak} $\lim_{n \rightarrow\infty} \sqrt{n}r_{n} = 0$ if
$\lim_{n \rightarrow
\infty} \sqrt{n}\|{\mathbf m} - {\mathbf m}^{(n)}\| = 0$ and $\lim_{n
\rightarrow\infty} n\sum_{k >
d(n)} \lambda_{k}^{2} = 0$. The latter is true for some $\alpha> 1/4$
if $\lim_{k \rightarrow
\infty} k^{2}\lambda_{k} = 0$ (e.g., if the $\lambda_{k}$'s decay
geometrically as $k
\rightarrow\infty$). We now discuss some conditions that are
sufficient to ensure $\lim_{n
\rightarrow\infty} \sqrt{n}\|{\mathbf m} - {\mathbf m}^{(n)}\| = 0$ as well as
$\lim_{n \rightarrow
\infty} \sqrt{n}s_{n} = 0$ [implying that $\lim_{n \rightarrow\infty}
\sqrt{n}\mbox{B}_{n}({\mathbf u}) = 0$] in separable Hilbert spaces. If
${\mathcal X} =
L_{2}([0,1],\nu)$, where $\nu$ is the uniform distribution, and $\{\phi
_{k}\}_{k \geq1}$ is
the set of standard Fourier basis functions, then Theorem $4.4$ in \cite
{Vret03} describes
those ${\mathbf x} \in{\mathcal X}$ for which $\lim_{n \rightarrow\infty}
\sqrt{n}\|{\mathbf x} - {\mathbf
x}^{(n)}\| = 0$ holds. It follows from that theorem that a sufficient
condition for $\lim_{n
\rightarrow\infty} \sqrt{n}\|{\mathbf x} - {\mathbf x}^{(n)}\| = 0$ to hold is
that ${\mathbf x}$ is
thrice differentiable on $[0,1]$, ${\mathbf x}(0) = {\mathbf x}(1)$, and its
right-hand side derivative at
$0$ equals its left-hand side derivative at $1$ for each of the three
derivatives. On the other
hand, if $\{\phi_{k}\}_{k \geq1}$ is either the set of normalized
Chebyshev or Legendre
polynomials, which form orthonormal bases of ${\mathcal X}$ when $\nu$
is the uniform and the
$\operatorname{Beta}(1/2,1/2)$ distributions, respectively, then ${\mathbf x} \in
{\mathcal X}$ satisfying $\lim_{n
\rightarrow\infty} \sqrt{n}\|{\mathbf x} - {\mathbf x}^{(n)}\| = 0$ can be
obtained using Theorem 4.2
in \cite{Tref08} and Theorem 2.1 in \cite{WX12}, respectively. Next,
let ${\mathcal X} =
L_{2}(\mathbb{R},\nu)$, where $\nu$ is the normal distribution with
zero mean and variance
$1/2$, and $\phi_{k}(t) \propto \exp\{-At^{2}\}h_{k}(A't), t \in\mathbb
{R}, k \geq1$ for an
appropriate $A \geq0$ and $A' > 0$, where $\{h_{k}\}_{k \geq1}$ is\vspace*{1pt}
the set of Hermite
polynomials. Then ${\mathbf x} \in{\mathcal X}$ satisfying $\lim_{n
\rightarrow\infty}
\sqrt{n}\|{\mathbf x} - {\mathbf x}^{(n)}\| = 0$ can be obtained from the
conditions of the theorem in
page~385 in \cite{Boyd84} for $j \geq5$. An important special case in
this setup is the
Gaussian process with the Gaussian covariance kernel, which is used in
classification and
regression problems (see, e.g., \cite{RW06}). The eigenvalues of this
kernel decay
geometrically, which implies that $\lim_{n \rightarrow\infty} n\sum_{k
> d(n)} \lambda_{k}^{2}
= 0$ for some $\alpha> 1/4$. Summarizing this discussion, we have the
following theorem.

\begin{teo} \label{teo7}
Suppose that the assumptions of Theorem~\ref{teo6} hold. Also, assume
that for some $\alpha\in
(1/4,1/2]$, $\sqrt{n}s_{n} \rightarrow0$, $\sqrt{n}\|{\mathbf m} - {\mathbf
m}^{(n)}\| \rightarrow0$
and $n\sum_{k > d(n)} \lambda_{k}^{2} \rightarrow0$ as $n \rightarrow
\infty$. Then, there
exists a zero mean Gaussian random element ${\mathbf Z}_{{\mathbf u}}$ such
that $\sqrt{n}
(\widehat{{\mathbf Q}}({\mathbf u}) - {\mathbf Q}({\mathbf u}))$ converges\vspace*{2pt} weakly to
${\mathbf Z}_{{\mathbf u}}$ as $n
\rightarrow\infty$. The covariance of ${\mathbf Z}_{{\mathbf u}}$ is given by
$V_{{\mathbf u}} =
[\widetilde{J}_{{\mathbf Q}({\mathbf u})}]^{-1}\Lambda_{{\mathbf u}}[\widetilde
{J}_{{\mathbf Q}({\mathbf
u})}]^{-1}$, where\vspace*{2pt} $\Lambda_{{\mathbf u}}\dvtx  {\mathcal X} \rightarrow
{\mathcal X}$ satisfies
$\langle\Lambda_{{\mathbf u}}({\mathbf z}),{\mathbf w}\rangle= E \{
\langle\frac{{\mathbf Q}({\mathbf u})
- {\mathbf X}}{\|{\mathbf Q}({\mathbf u}) - {\mathbf X}\|} - {\mathbf u},{\mathbf z}
\rangle \langle
\frac{{\mathbf Q}({\mathbf u}) - {\mathbf X}}{\|{\mathbf Q}({\mathbf u}) - {\mathbf X}\|} -
{\mathbf u},{\mathbf
w} \rangle \}$ for ${\mathbf z}$, ${\mathbf w} \in{\mathcal X}$,\vspace*{2pt} and
$\langle\cdot,\cdot\rangle$ denotes
the inner product in ${\mathcal X}$.
\end{teo}
A random element ${\mathbf Z}$ in the separable Hilbert space ${\mathcal
X}$ is said to have a Gaussian distribution with mean ${\mathbf m} \in
{\mathcal X}$ and covariance ${\mathbf C}\dvtx  {\mathcal X} \rightarrow
{\mathcal X}$ if for any ${\mathbf l} \in{\mathcal X}$, $\langle{\mathbf
l},{\mathbf Z}\rangle$ has a Gaussian distribution on $\mathbb{R}$ with
mean $\langle{\mathbf l},{\mathbf m}\rangle$ and variance $\langle{\mathbf C}({\mathbf
l}),{\mathbf l}\rangle= E\{(\langle{\mathbf l},{\mathbf Z} - {\mathbf m}\rangle)^{2}\}
$ (see, e.g., \cite{AG80}).

\subsection{Asymptotic efficiency of the sample spatial median} \label{subsec43}

We will now study the asymptotic efficiency of the sample
spatial median $\widehat{{\mathbf Q}}({\mathbf0})$ relative to the sample mean
$\overline{{\mathbf X}}$ when ${\mathbf X}$\vadjust{\goodbreak} has a symmetric distribution in a
Hilbert space ${\mathcal X}$ about some ${\mathbf m} \in{\mathcal X}$.
In this case, ${\mathbf Q}({\mathbf0}) = E({\mathbf X}) = {\mathbf m}$. We assume that
$E(\|{\mathbf X}\|^{2}) < \infty$, and let $\Sigma$ be the covariance of
${\mathbf X}$. Note that ${\mathbf Q}_{n}({\mathbf0}) = {\mathbf m}^{(n)}$, and
following the discussion after Theorem~\ref{teo6}, it can be shown that
under the conditions of that theorem and if $\sqrt{n}\|{\mathbf m} - {\mathbf
m}^{(n)}\| \rightarrow0$ as $n \rightarrow\infty$, we have the weak
convergence of $\sqrt{n}(\widehat{{\mathbf Q}}({\mathbf0}) - {\mathbf m})$ to
${\mathbf Z}_{{\mathbf0}}$ as $n \rightarrow\infty$. Here, ${\mathbf Z}_{{\mathbf
0}}$ is a Gaussian random element with zero mean and covariance
$V_{{\mathbf0}}$ as in Theorem~\ref{teo7}. On the other hand, using the
central limit theorem in Hilbert spaces, we have the weak convergence
of $\sqrt{n}(\overline{{\mathbf X}} - {\mathbf m})$ to a Gaussian random
element with zero mean and covariance $\Sigma$.

 For our asymptotic efficiency study, we have first considered
${\mathbf X} = {\mathbf m} +
\sum_{k=1}^{\infty} \lambda_{k}Y_{k}\phi_{k}$ in $L_{2}[0,1]$ with
$Y_{k}$'s having independent
standard normal distributions, and the $\lambda_{k}^{2}$'s and the $\phi
_{k}$'s being the
eigenvalues and the eigenfunctions of the covariance kernel $K(t,s) =
0.5(t^{2{\rm{H}}} +
s^{2{\rm{H}}} - |t-s|^{2{\rm{H}}})$ for $\mbox{H}$ ranging from $0.1$
to $0.9$. In this case,
${\mathbf X}$ has the distribution of a fractional Brownian motion on
$[0,1]$ with mean ${\mathbf m}$
and Hurst index $\mbox{H}$. We have also considered $t$-processes (see,
e.g., \cite{YTY07}) on
$[0,1]$ with mean ${\mathbf m}$, degrees of freedom $r \geq3$ and
covariance kernel $K(t,s) =
\min(t,s)$. In this case, ${\mathbf X} = {\mathbf m} + \sum_{k=1}^{\infty}
\lambda_{k}Y_{k}\phi_{k}$
with $Y_{k} = Z_{k}/\sqrt{W/r}$ for $r \geq3$, where the $Z_{k}$'s are
independent standard
normal variables, and $W$ is an independent chi-square variable with
$r$ degrees of freedom.
Here, the $\lambda_{k}^{2}$'s and the $\phi_{k}$'s are the eigenvalues
and the eigenfunctions,
respectively, of the covariance kernel $K(t,s) = \min(t,s)$. We have
also included in our study
the distributions of ${\mathbf X} = {\mathbf m} + \sum_{k=1}^{\infty} \lambda
_{k}Y_{k}\phi_{k}$ in
$L_{2}(\mathbb{R},\nu)$ corresponding to all the choices of the
$Y_{k}$'s mentioned above.
Here, $\nu$ is the normal distribution with zero mean and variance
$1/2$, the
$\lambda_{k}^{2}$'s and the $\phi_{k}$'s are the eigenvalues and the
eigenfunctions,
respectively, of the Gaussian covariance kernel $K(t,s) = \exp\{
-(t-s)^{2}\}$ (see Section~4.3
in \cite{RW06}). These processes on $\mathbb{R}$ are the Gaussian and
the $t$-processes with $r$
degrees of freedom for $r \geq3$, respectively, having mean ${\mathbf m}$
and the Gaussian
covariance kernel. The mean function ${\mathbf m}$ of each of the processes
considered above is
assumed to satisfy $\sqrt{n}\|{\mathbf m} - {\mathbf m}^{(n)}\| \rightarrow0$
as $n \rightarrow
\infty$ so that\vspace*{1pt} we can apply Theorem~\ref{teo7}. The asymptotic
efficiency of $\widehat{{\mathbf
Q}}({\mathbf0})$ relative to $\overline{{\mathbf X}}$ can be defined as
$\trace(\Sigma)/\trace(V_{{\mathbf
0}})$. The traces of $\Sigma$ and $V_{{\mathbf0}}$ are defined as $\sum_{k=1}^{\infty} \langle
\Sigma\psi_{k},\psi_{k}\rangle$ and $\sum_{k=1}^{\infty} \langle V_{{\mathbf
0}}\psi_{k},\psi_{k}\rangle$, respectively, where $\{\psi_{k}\}_{k \geq
1}$ is an orthonormal
basis of the Hilbert space ${\mathcal X}$. It can be shown that both
the infinite sums are
convergent, and their values are independent of the choice of $\{\psi
_{k}\}_{k \geq1}$. For
numerically computing the efficiency, each of the two infinite
dimensional covariances are
replaced by the $D$-dimensional covariance matrix of the distribution
of $({\mathbf X}(t_{1}),{\mathbf
X}(t_{2}),\ldots,{\mathbf X}(t_{D}))$, where $D$ is appropriately large.
For the processes in
$L_{2}[0,1]$, $t_{1},t_{2},\ldots,t_{D}$ are chosen to be equispaced
points in $[0,1]$, while
for the processes in $L_{2}(\mathbb{R},\nu)$, these points are chosen
randomly from the
distribution $\nu$. These choices ensure that for any ${\mathbf x} \in
L_{2}[0,1]$ or
$L_{2}(\mathbb{R},\nu)$, $\|{\mathbf x}\|^{2}$ can be approximated by the
average of ${\mathbf
x}^{2}(t)$ over these $D$ points. For our numerical evaluation of the
asymptotic efficiencies,
we have chosen $D = 200$.

 The efficiency of $\widehat{{\mathbf Q}}({\mathbf0})$ relative to
$\overline{{\mathbf X}}$ for the fractional Brownian motion decreases from
$0.923$ to $0.718$ as the value of $\mbox{H}$ increases from $0.1$ to
$0.9$. For the Brownian motion (i.e., when $H = 0.5$) this efficiency
is $0.83$. For the $t$-processes in $[0,1]$, this efficiency is $2.135$
for $3$ degrees of freedom, and it decreases with the increase in the
degrees of freedom. The efficiency remains more than $1$ up to $9$
degrees of freedom, when its value is $1.006$. This efficiency for the
Gaussian process in $L_{2}(\mathbb{R},\nu)$ is $0.834$. The efficiency
for the $t$-processes in $L_{2}(\mathbb{R},\nu)$ is $2.247$ for $3$
degrees of freedom, and it decreases with the increase in the degrees
of freedom. As before, this efficiency remains more than $1$ up to $9$
degrees of freedom, when its value is $1.013$.

\section{Spatial depth and the DD-plot in Banach spaces}\label{sec3}

In the finite dimensional setup, the spatial distribution has
been used to define the spatial depth (see \cite{Serf02} and \cite
{VZ00}). Likewise, the spatial depth at ${\mathbf x}$ in a smooth Banach
space ${\mathcal X}$ with respect to the probability distribution of a
random element ${\mathbf X} \in{\mathcal X}$ can be defined as $\SD({\mathbf
x}) = 1 - \|S_{{\mathbf x}}\|$, and its empirical version is given by
$\widehat{\SD}({\mathbf x}) = 1 - \|\widehat{S}_{{\mathbf x}}\|$. Here, $S_{{\mathbf
x}}$ and $\widehat{S}_{{\mathbf x}}$ are as defined in Section~\ref{sec2}.
There are a few other notions of depth function for data in infinite
dimensional function spaces (see, e.g., \cite{FM01,LPR09,LPR11} and \cite{SG11}). However, as shown in \cite{CC14}, some of
these depth functions exhibit degeneracy for certain types of
functional data, and hence are not very useful.

 We will now discuss some properties of the spatial depth
function in Banach spaces. The spatial distribution function $S_{{\mathbf
x}}$ possesses an invariance property under the class of affine
transformations $L\dvtx  {\mathcal X} \rightarrow{\mathcal X}$ of the form
$L({\mathbf x}) = cA({\mathbf x}) + {\mathbf a}$, where $c > 0$, ${\mathbf a} \in
{\mathcal X}$ and $A\dvtx  {\mathcal X} \rightarrow{\mathcal X}$ is a
linear surjective isometry. By the definition of G\^ateaux derivative
and using the isometry of $A$, we have
\begin{eqnarray*}
\SGN_{L({\mathbf x}) - L({\mathbf X})}({\mathbf h}) &=& \SGN_{A({\mathbf x}) - A({\mathbf X})}\bigl(A\bigl({\mathbf h}'\bigr)\bigr)
=  \SGN_{{\mathbf
x} - {\mathbf X}}\bigl({\mathbf h}'\bigr)
\\
&=& \SGN_{{\mathbf x} - {\mathbf X}}\bigl(A^{-1}({\mathbf h})\bigr)
=
\bigl(A^{-1}\bigr)^{*}\bigl(\SGN_{{\mathbf x} - {\mathbf X}}({\mathbf h})\bigr)
\end{eqnarray*}
for\vspace*{1pt} any ${\mathbf x}, {\mathbf
h} \in{\mathcal X}$. Here, ${\mathbf h} = A({\mathbf h}')$, and $(A^{-1})^{*}\dvtx  {\mathcal X}^{*} \rightarrow{\mathcal X}^{*}$ denotes the adjoint of~$A^{-1}$. Thus, if $S_{L({\mathbf x})}$ is the spatial distribution at
$L({\mathbf x})$ with respect to the probability distribution of $L({\mathbf
X})$, we have $S_{L({\mathbf x}
)} = (A^{-1})^{*}(S_{{\mathbf x}})$, where $S_{{\mathbf x}}$ is the spatial
distribution at ${\mathbf x}$ with respect to the probability distribution
of ${\mathbf X}$. This implies that the spatial depth is invariant under
such affine transformations in the sense that the spatial depth at
$L({\mathbf x})$ with respect to the distribution of $L({\mathbf X})$ is the
same as the spatial depth at ${\mathbf x}$ with respect to the distribution
of ${\mathbf X}$.

 It follows from Remark 3.5 and Theorems 2.17 and 4.14 in \cite
{Kemp87} that if ${\mathcal X}$ is a strictly convex Banach space, and
the distribution of ${\mathbf X}$ is nonatomic and not entirely contained
on a line in ${\mathcal X}$, then $\SD({\mathbf x})$ has a unique maximizer
at the spatial median (say, ${\mathbf m}$) of ${\mathbf X}$ and $\SD({\mathbf m}) =
1$. It follows from the last assertion in Theorem~\ref{teo1} that if
the norm in ${\mathcal X}$ is Fr\'echet differentiable and the
distribution of ${\mathbf X}$ is nonatomic, then $\SD({\mathbf x})$ is a
continuous function in ${\mathbf x}$. Moreover, in such cases, $\SD({\mathbf x}
+ n{\mathbf y}) \rightarrow0$ as $n \rightarrow\infty$ for any ${\mathbf x},
{\mathbf y} \in{\mathcal X}$ with ${\mathbf y} \neq{\mathbf0}$. This implies
that the spatial depth function vanishes at infinity along any ray
through any point in ${\mathcal X}$. The above properties of $\SD({\mathbf
x})$ are among the desirable properties of any statistical depth
function listed in \cite{Liu90} and \cite{ZS00a} for the finite dimensional
setting.

 It follows from Theorem~\ref{teo1} that if ${\mathcal X}$ is a
reflexive Banach space and the distribution of ${\mathbf X}$ is nonatomic,
then $\SD({\mathbf x})$ takes all values in $(0,1]$ as ${\mathbf x}$ varies over
${\mathcal X}$. Also, if $\SD({\mathbf x})$ is continuous in ${\mathbf x}$, then
$\SD({\mathbf x})$ takes all values in $(0,w] \subseteq(0,1]$ as ${\mathbf x}$
varies over a closed subspace ${\mathcal W}$ of ${\mathcal X}$, where
$w = \sup_{{\mathbf x} \in{\mathcal W}} \SD({\mathbf x})$. In particular, $w =
1$ if ${\mathcal W}$ contains the spatial median of ${\mathbf X}$. It can
be shown that the support of a Gaussian distribution in a separable
Banach space is the closure of the translation of a subspace of
${\mathcal X}$ by the mean (which is also the spatial median) of that
distribution. So, if the norm in that space is Fr\'echet
differentiable, then $\SD({\mathbf x})$ is continuous in ${\mathbf x}$ and it
takes all values in $(0,1]$ as ${\mathbf x}$ varies over the support of
that distribution.

 The properties of the spatial depth discussed above imply that
it induces a meaningful center-outward ordering of the points in these
spaces, and can be used to develop depth-based statistical procedures
for data from such distributions. On the other hand, many of the
well-known depths for infinite dimensional data like the half-space
depth, the band depth and the half-region depth do not possess such
regular behavior and exhibit degeneracy for many Gaussian distributions
(see \cite{CC14}).

 We will next study the properties of the empirical spatial
depth in smooth Banach spaces. A Banach space ${\mathcal X}$ is said to
be of type $2$ (see, e.g., \cite{AG80}) if there exists a constant
$\gamma> 0$ such that for any $n\geq1$ and independent zero mean
random elements ${\mathbf U}_{1},{\mathbf U}_{2},\ldots,{\mathbf U}_{n}$ in
${\mathcal X}$ with $E\{\|{\mathbf U}_{i}\|^{2}\} < \infty$ for all
$i=1,2,\ldots,n$, we have $E\{\|\sum_{i=1}^{n} {\mathbf U}_{i}\|^{2}\} \leq
{\gamma}\sum_{i=1}^{n} E\{\|{\mathbf U}_{i}\|^{2}\}$. Examples of type $2$
spaces include Hilbert spaces and $L_{p}$ spaces with $p \geq2$. Type
$2$ Banach spaces are the only Banach spaces, where the central limit
theorem will hold for every sequence of i.i.d. random elements, whose
squared norms have finite expectations. Let ${\mathbf C}\dvtx  {\mathcal X}^{*}
\rightarrow{\mathcal X}^{**}$ be a symmetric nonnegative definite
continuous linear operator. A random element ${\mathbf X}$ in a separable
Banach space ${\mathcal X}$ is said to have a Gaussian distribution
with mean\vadjust{\goodbreak} ${
\mathbf m} \in{\mathcal X}$ and covariance ${\mathbf C}$ if for any ${\mathbf l}
\in{\mathcal X}^{*}$, ${\mathbf l}({\mathbf X})$ has a Gaussian distribution
on $\mathbb{R}$ with mean ${\mathbf l}({\mathbf m})$ and variance $({\mathbf
C}({\mathbf l}))({\mathbf l})$ (see, e.g., \cite{AG80}). If ${\mathcal X}$ is a
Hilbert space, this definition coincides with the one given after
Theorem~\ref{teo7}.
%
\begin{teo} \label{teo4}
Suppose that the assumptions of part~\textup{(a)} of Theorem~\ref{teo2} hold.
Then, $\sup_{{\mathbf x} \in K} |\widehat{\SD}({\mathbf x}) - \SD({\mathbf x})|
\rightarrow0$ as $n \rightarrow\infty$ \emph{almost surely} for every
compact set $K \subseteq{\mathcal X}$. Suppose that the norm function
in ${\mathcal X}^{*}$ is Fr\'echet differentiable, and ${\mathcal
X}^{*}$ is a separable and type $2$ Banach space. Then $\sqrt
{n}(\widehat{\SD}({\mathbf x}) - \SD({\mathbf x}))$ converges weakly to
$\SGN_{S_{{\mathbf x}}}({\mathbf W})$ if $S_{{\mathbf x}} \neq{\mathbf0}$. If $S_{{\mathbf
x}} = {\mathbf0}$, $\sqrt{n}(\widehat{\SD}({\mathbf x}) - \SD({\mathbf x}))$
converges weakly to $-\|{\mathbf V}\|$. Here, ${\mathbf W}$ and ${\mathbf V}$ are
zero mean Gaussian random elements in ${\mathcal X}^{*}$.
\end{teo}

In the finite dimensional setup, an exploratory data analytic
tool for checking whether two given samples arise from the same
distribution or not is the depth--depth plot (DD-plot) (see \cite
{LPS99}). A DD-plot is a scatter plot of the depth values of the data
points in the pooled sample with respect to the empirical distributions
of the two samples. It can be used to detect differences in location,
scale, etc. Here, we consider the problem of constructing DD-plots for
data in infinite dimensional spaces. It follows from \cite{CC14} that
the half-space depth and the simplicial depth, which have been used by
the authors of \cite{LPS99} for constructing DD-plots for data in
finite dimensional spaces, cannot be used for constructing DD-plots in
infinite dimensional spaces.

\begin{figure}

\includegraphics{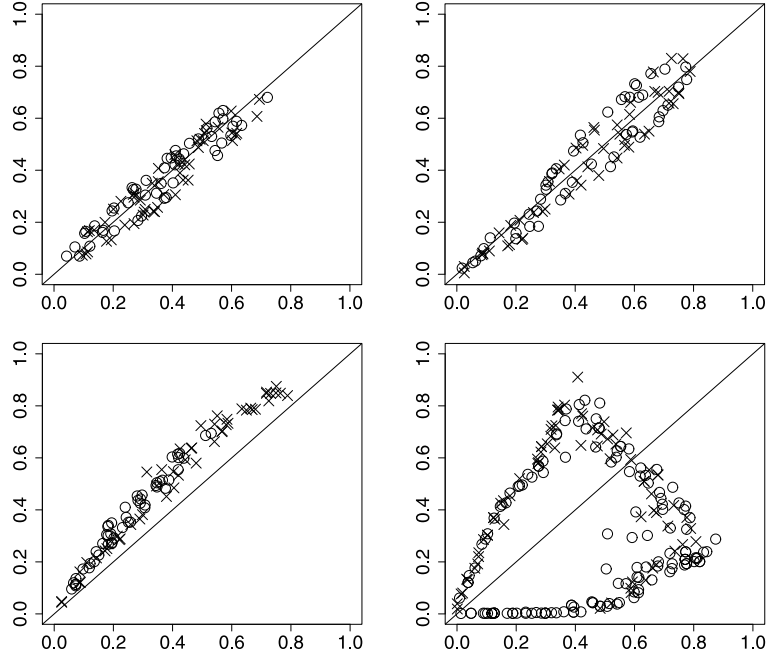}

\caption{The DD-plots for the simulated and the real data. The first
(resp., second) figure from the left is the DD-plot for the two samples
from the standard Brownian motion (resp., the fractional
Brownian motion). The third figure is the DD-plot for the
two samples from the standard Brownian motion and the fractional
Brownian motion. The fourth figure is the DD-plot for the two samples
in the spectrometric data.}
\label{Fig4}
\end{figure}

 We have prepared DD-plots for some real and simulated
functional data using the spatial depth (see Figure~\ref{Fig4}). The
simulated datasets are samples from the standard Brownian motion and
the fractional Brownian motion with ${\rm{H}} = 0.9$. Both of these
processes have Karhunen--Lo\`eve expansions in $L_{2}[0,1]$ (see
Section~\ref{sec4}). Each simulated data consists of $n = 50$ samples,
and the sample curves are observed at $250$ equispaced points on
$[0,1]$. The real data is the spectrometry data used in Section~\ref
{sec4}, which can be viewed as a random sample from a probability
distribution in $L_{2}[850,1050]$. Since the sample spaces for the
simulated and the real datasets considered here are Hilbert spaces,
$S_{{\mathbf x}}$ simplifies to $E\{({\mathbf x} - {\mathbf X})/\|{\mathbf x} - {\mathbf
X}\|\}$. The norm in this expression is computed as the norm of the
Euclidean space whose dimension is the number of values of the argument
over which the sample functions in the dataset are observed.

%

 The first (resp., second) plot in Figure~\ref{Fig4} is
the DD-plot for the two
samples from the standard Brownian motion (resp., the fractional
Brownian motion). The
third plot is the DD-plot for the two samples from the standard
Brownian motion and the
fractional Brownian motion. The axes of the first and the second
DD-plots correspond to the
depth values with respect to the empirical distributions of the
standard Brownian motion and
the fractional Brownian motion, respectively. In each of those plots,
the $\circ$'s and the
$\times$'s represent the sample observations of the two samples. The
vertical and the
horizontal axis of the third DD-plot correspond to the depth values
with respect to the
empirical distributions of the standard Brownian motion and the
fractional Brownian motion,
respectively, and the $\circ$'s and the $\times$'s represent the
samples from these two
distributions, respectively. In the first two DD-plots, the $\circ$'s
and the $\times$'s are
clustered around the $45^{\circ}$ line through the origin. So, the
observations from each of
the two samples have similar depth values with respect to both the
samples. This indicates that
there is not much difference between the two underlying populations in
each case. In the third
DD-plot, all the $\circ$'s and the $\times$'s lie above the $45^{\circ
}$ line through the
origin in the shape of an arch. So, all the observations in the sample
from the fractional
Brownian motion have higher depth values with respect to the empirical
distribution of the
sample from the standard Brownian motion. This indicates that the
former population has less
spread than the latter one. The horizontal and the vertical axes of the
DD-plot for the
spectrometric data (see the fourth plot in Figure~\ref{Fig4})
correspond to the spatial depth
values with respect to the empirical distribution of the classes with
fat content $\leq$20\%
and $>$20\%, respectively, and the $\circ$'s and the $\times$'s
represent the samples from
these two classes, respectively. It is seen that the observations from
both the samples are
almost evenly spread out below and above the $45^{\circ}$ line through
the origin in the shape
of a triangle. One side of the triangle is formed by the line joining
the points with
approximate coordinates $(0.4,0.8)$ and $(0.8,0.2)$, and the vertex
opposite to that side is
the origin. This type of DD-plot indicates a difference in location
between the two samples.
The points around the aforementioned side of the triangle lie in the
overlapping region of the
two samples, and have moderate to high depth values with respect to the
empirical distributions
of both the samples.

\begin{appendix}\label{sec5}
\section*{Appendix: The proofs}

The proofs involve several concepts and techniques from probability
theory in Banach spaces and convex analysis. Readers are referred to
\cite{AG80} for an exposition on probability theory in Banach spaces.
We refer to \cite{FHH01} for an exposition on the theory of Banach
spaces, and \cite{BV10} for the relevant details on convex analysis in
Banach spaces.
%
\begin{lemma} \label{lemma0}
Suppose that ${\mathcal X}^{*}$ is a separable Banach space. If $\mu$
is atomic, then $\sup_{{\mathbf x} \in{\mathcal X}} \|\widehat{S}_{{\mathbf
x}} - S_{{\mathbf x}}\| \rightarrow0$ as $n \rightarrow\infty$ \emph{almost surely}.
\end{lemma}

\begin{pf}
Define $\widehat{p}({\mathbf y}) = n^{-1} \sum_{i=1}^{n} I({\mathbf X}_{i} =
{\mathbf y})$ and $p({\mathbf y}) = P({\mathbf X} = {\mathbf y})$ for \mbox{${\mathbf y} \in
A_{\mu}$}, where $A_{\mu}$ denotes the set of atoms of $\mu$. By the
strong law of large numbers, $\lim_{n \rightarrow\infty} \widehat
{p}({\mathbf y}) = p({\mathbf y})$ \emph{almost surely} for each ${\mathbf y} \in
A_{\mu}$. Observe that $\sup_{{\mathbf x} \in{\mathcal X}} \|\widehat
{S}_{{\mathbf x}} - S_{{\mathbf x}}\| \leq\sum_{{\mathbf y} \in A_{\mu}} |\widehat
{p}({\mathbf y}) - p({\mathbf y})| = 2 - 2\sum_{{\mathbf y} \in A_{\mu}} \min\{
\widehat{p}({\mathbf y}),p({\mathbf y})\}$.\vspace*{1pt} Since $\min\{\widehat{p}({\mathbf
y}),p({\mathbf y})\} \leq p({\mathbf y})$, the proof is complete by the
dominated convergence theorem.
\end{pf}

\begin{pf*}{Proof of Theorem~\ref{teo}}
Let us write $\mu= \rho\mu_{1} + (1-\rho)\mu_{2}$, where $\mu_{1}$ and
$\mu_{2}$ are the nonatomic and the atomic parts of $\mu$,
respectively. Let $N_{n} = \sum_{i=1}^{n} I({\mathbf X}_{i} \notin A_{\mu
})$, where $A_{\mu}$ is the set of atoms of $\mu$. Denote by $\widehat
{\mu}_{1}$ and $\widehat{\mu}_{2}$ the empirical probability
distributions corresponding to $\mu_{1}$ and $\mu_{2}$, respectively.
Here, as well as in other proofs in this section, we will denote the
inner product in a Hilbert space by $\langle\cdot,\cdot\rangle$. Observe that
for any ${\mathbf x} \in{\mathcal Z}$ and ${\mathbf l} \in{\mathcal X}$,
\begin{eqnarray*}
\bigl\llvert \langle{\mathbf l},\widehat{S}_{{\mathbf x}} - S_{{\mathbf x}}
\rangle\bigr\rrvert &\leq& \biggl\llvert \frac{N_{n}}{n}E_{\widehat{\mu}_{1}} \biggl(
\biggl\langle{\mathbf l},\frac{{\mathbf x} - {\mathbf X}}{\|{\mathbf x} - {\mathbf X}\|
} \biggr\rangle \biggr) -
\frac{N_{n}}{n}E_{\mu_{1}} \biggl( \biggl\langle {\mathbf l},\frac{{\mathbf x} - {\mathbf X}}{\|{\mathbf x} - {\mathbf X}\|}
\biggr\rangle \biggr)\biggr\rrvert
\\
&&{} + \biggl\llvert \frac{N_{n}}{n}E_{\mu_{1}} \biggl( \biggl\langle{\mathbf
l},\frac{{\mathbf x} - {\mathbf X}}{\|{\mathbf x} - {\mathbf X}\|} \biggr\rangle \biggr) - {\rho}E_{\mu_{1}} \biggl(
\biggl\langle{\mathbf l},\frac{{\mathbf x} - {\mathbf X}}{\|
{\mathbf x} - {\mathbf X}\|} \biggr\rangle \biggr)\biggr\rrvert
\\
&&{} + \biggl\llvert \frac{n-N_{n}}{n}E_{\widehat{\mu}_{2}} \biggl( \biggl\langle{\mathbf
l},\frac{{\mathbf x} - {\mathbf X}}{\|{\mathbf x} - {\mathbf X}\|} \biggr\rangle \biggr) - \frac{n-N_{n}}{n}E_{\mu_{2}}
\biggl( \biggl\langle{\mathbf l},\frac{{\mathbf x} - {\mathbf X}}{\|{\mathbf x} - {\mathbf X}\|} \biggr\rangle \biggr)\biggr\rrvert
\\
&&{} + \biggl\llvert \frac{n-N_{n}}{n}E_{\mu_{2}} \biggl( \biggl\langle{\mathbf
l},\frac{{\mathbf x} - {\mathbf X}}{\|{\mathbf x} - {\mathbf X}\|} \biggr\rangle \biggr) - (1-\rho)E_{\mu_{2}} \biggl(
\biggl\langle{\mathbf l},\frac{{\mathbf x} - {\mathbf
X}}{\|{\mathbf x} - {\mathbf X}\|} \biggr\rangle \biggr)\biggr\rrvert
\\
&\leq& \biggl\llvert \biggl\langle{\mathbf l},E_{\widehat{\mu}_{1}} \biggl(
\frac
{{\mathbf x} - {\mathbf X}}{\|{\mathbf x} - {\mathbf X}\|} \biggr) - E_{\mu_{1}} \biggl(\frac{{\mathbf x} - {\mathbf X}}{\|{\mathbf x} - {\mathbf X}\|} \biggr)
\biggr\rangle \biggr\rrvert
\\
&&{} + \biggl\llvert \biggl\langle{\mathbf l},E_{\widehat{\mu}_{2}} \biggl(
\frac
{{\mathbf x} - {\mathbf X}}{\|{\mathbf x} - {\mathbf X}\|} \biggr) - E_{\mu_{2}} \biggl(\frac{{\mathbf x} - {\mathbf X}}{\|{\mathbf x} - {\mathbf X}\|} \biggr)
\biggr\rangle \biggr\rrvert + 2\llvert N_{n}/n - \rho\rrvert.
\end{eqnarray*}
In other words,
%
%
\begin{eqnarray}\label{eq211a}
\bigl\llvert \langle{\mathbf l},\widehat{S}_{{\mathbf x}} - S_{{\mathbf x}}
\rangle\bigr\rrvert &\leq&\biggl\llvert \biggl\langle{\mathbf l},E_{\widehat{\mu}_{1}}
\biggl(\frac{{\mathbf x} - {\mathbf X}}{\|{\mathbf x} - {\mathbf X}\|} \biggr) - E_{\mu
_{1}} \biggl(\frac{{\mathbf x} - {\mathbf X}}{\|{\mathbf x} - {\mathbf X}\|}
\biggr) \biggr\rangle\biggr\rrvert\nonumber
\\
&&{}  +
\|{\mathbf l}\| \biggl\llVert E_{\widehat{\mu}_{2}} \biggl(\frac{{\mathbf x} - {\mathbf
X}}{\|{\mathbf x} - {\mathbf X}\|} \biggr) -
E_{\mu_{2}} \biggl(\frac{{\mathbf x} -
{\mathbf X}}{\|{\mathbf x} - {\mathbf X}\|} \biggr)\biggr\rrVert
\\
&&{} + 2\llvert N_{n}/n - \rho \rrvert.\nonumber
\end{eqnarray}
The third term in the right-hand side of (\ref{eq211a}) converges to
zero as $n \rightarrow\infty$ \emph{almost surely} by the strong law of
large numbers. By Lemma~\ref{lemma0}, the second term in the right-hand
side of (\ref{eq211a}) converges to zero \emph{uniformly over} ${\mathbf
x} \in{\mathcal X}$ as $n \rightarrow\infty$ \emph{almost surely}.

 Let us next consider the class of functions
\begin{eqnarray*}
{\mathcal G} &=& \bigl\{ \psi_{{\mathbf x}}\dvtx  {\mathcal X} \rightarrow\mathbb{R},
\psi_{{\mathbf x}}({\mathbf s}) = \langle{\mathbf l},{\mathbf x} - {\mathbf s}\rangle I({\mathbf x}
\neq{\mathbf s})/\|{\mathbf x} - {\mathbf s}\|; {\mathbf x} \in{\mathcal Z} \bigr\}.
\end{eqnarray*}
Similar arguments as those in the proofs of Theorems 5.5 and 5.6 in pages
471--474 in \cite{Kolt97} show that ${\mathcal G}$ is a VC-subgraph
class. Since $\mu_{1}$ is nonatomic, the functions in ${\mathcal G}$
are \emph{almost surely} $\mu_{1}$-continuous. Thus, using the
separability of ${\mathcal X}$, we get that ${\mathcal G}$ is a
point-wise separable class (see page~116 in~\cite{VW96}) with an envelope
function that is unity everywhere. Thus, it follows from Theorem
$2.6.8$ in \cite{VW96} that ${\mathcal G}$ is a Glivenko--Cantelli
class with respect to the measure $\mu_{1}$, which implies that the
first term in the right-hand side of (\ref{eq211a}) converges \emph{uniformly over} ${\mathbf x} \in{\mathcal Z}$ as $n \rightarrow\infty$
\emph{almost surely}.

 Since ${\mathcal X}$ is separable, it has a countable dense
subset ${\mathcal L}$. So,
%
%
\begin{eqnarray}\qquad
&& \lim_{n \rightarrow\infty} \sup_{{\mathbf x} \in{\mathcal Z}} \biggl\llvert
\biggl\langle{\mathbf l},E_{\widehat{\mu}_{1}} \biggl(\frac{{\mathbf x} - {\mathbf
X}}{\|{\mathbf x} - {\mathbf X}\|} \biggr) -
E_{\mu_{1}} \biggl(\frac{{\mathbf x} -
{\mathbf X}}{\|{\mathbf x} - {\mathbf X}\|} \biggr) \biggr\rangle\biggr\rrvert = 0\qquad
\forall {\mathbf l} \in{\mathcal L} \label{eq211b}
\end{eqnarray}
as $n \rightarrow\infty$ \emph{almost surely}. Note that both the
expectations in (\ref{eq211b}) above are bounded in norm by $1$.
Using this fact, equation (\ref{eq211b}) and the fact that ${\mathcal
L}$ is dense in ${\mathcal X}$, we get the proof.

 For the second part of the theorem, note that it is enough to
prove the result for $d = 1$. By the Riesz representation theorem, for
any continuous linear map \mbox{${\mathbf g}\dvtx  {\mathcal X} \rightarrow\mathbb
{R}$}, there exists ${\mathbf l} \in{\mathcal X}$ satisfying ${\mathbf g}({\mathbf
x}) = \langle{\mathbf l},{\mathbf x}\rangle$ for every ${\mathbf x} \in{\mathcal
X}$. Let us consider the class of functions ${\mathcal G}$ defined
above in the proof of the first part of this theorem. If $\mu$ itself
is nonatomic, it follows from the arguments in that proof by replacing
$\mu_{1}$ with $\mu$ that ${\mathcal G}$ is a VC-subgraph class. This
along with Theorem 2.6.8 in \cite{VW96} implies that ${\mathcal G}$ is
a Donsker class with respect to $\mu$. This completes the proof of the theorem.
\end{pf*}

\begin{rem*}
Suppose that ${\mathcal X} = L_{p}$ for an even
integer $p > 2$. Using arguments similar to those used in deriving (\ref
{eq211a}), we get an analogous bound for ${\mathbf l}(\SGN_{{\mathbf x} - {\mathbf
X}})$ for any ${\mathbf x} \in{\mathcal Z}$ and ${\mathbf l} \in{\mathcal
X}$. In this case, ${\mathcal G}$ in the proof of Theorem~\ref{teo} is
to be defined as ${\mathcal G} = \{ \psi_{{\mathbf x}}\dvtx  {\mathcal X}
\rightarrow\mathbb{R},  \psi_{{\mathbf x}}({\mathbf s}) = {\mathbf l}(\SGN_{{\mathbf
x} - {\mathbf s}}); {\mathbf x} \in{\mathcal Z} \}$, and ${\mathbf g}$ in that
theorem is to be chosen a function from ${\mathcal X}^{*}$ into $\mathbb
{R}^{d}$. Using arguments similar to those in the proof of Theorem~\ref
{teo}, it can be shown that ${\mathcal G}$ is a VC-subgraph and a
point-wise separable class, and hence a Glivenko--Cantelli and a
Donsker class. So, the assertions of Theorem~\ref{teo} hold in this
case as well.
\end{rem*}

The following fact is a generalization of the Bernstein
inequality for probability distributions in separable Hilbert spaces,
and it will be used in the proof of Theorem~\ref{teo2}(b).
%
\begin{fact}[(\cite{Yuri76}, page 491)] \label{fact1}
Let ${\mathbf Y}_{1}, {\mathbf Y}_{2}, \ldots, {\mathbf Y}_{n}$ be independent
random elements in a separable Hilbert space ${\mathcal X}$ satisfying
$E({\mathbf Y}_{i}) = {\mathbf0}$ for $1 \leq i \leq n$. Suppose that for some
$h > 0$ and $u_{i} > 0$, we have $E(\|{\mathbf Y}_{i}\|^{m}) \leq
(m!/2)u_{i}^{2}h^{m-2}$ for $1 \leq i \leq n$ and all $m \geq2$. Let
$U_{n}^{2} = \sum_{i=1}^{n} u_{i}^{2}$. Then, for any $v > 0$, $P(\|\sum_{i=1}^{n} {\mathbf Y}_{i}\| \geq vU_{n}) \leq2\exp\{-(v^{2}/2)(1 +
1.62(vh/U_{n}))^{-1}\}$.
\end{fact}

\begin{pf*}{Proof of Theorem~\ref{teo2}}
(a) As in the proof of Theorem~\ref{teo}, we get
%
%
\begin{eqnarray}\label{eq1}\qquad
\|\widehat{S}_{{\mathbf x}} - S_{{\mathbf x}}\| &\leq& \bigl\|E_{\widehat{\mu}_{1}}
\{\SGN_{{\mathbf x} - {\mathbf X}}\} - E_{\mu
_{1}}\{\SGN_{{\mathbf x} - {\mathbf X}}\}\bigr\|
\nonumber\\[-8pt]\\[-8pt]
&&{} +
\bigl\|E_{\widehat{\mu}_{2}} \{\SGN_{{\mathbf x} - {\mathbf X}}\} - E_{\mu_{2}} \{
\SGN_{{\mathbf x} - {\mathbf X}}\}\bigr\| + 2\llvert N_{n}/n - \rho\rrvert.
\nonumber
\end{eqnarray}
Further, the second and the third terms in the right-hand side of the
inequality in~(\ref{eq1}) converge to zero as $n \rightarrow\infty$
\emph{almost surely} by the same arguments as in the proof of Theorem
\ref{teo}. Note that the convergence of the second term is uniform in~${\mathcal X}$ as before.

 Now, for an $\varepsilon> 0$, consider an $\varepsilon$-net
${\mathbf v}_{1}, {\mathbf v}_{2}, \ldots, {\mathbf v}_{N(\varepsilon)}$ of $K$.
The first term in the right-hand side of the inequality in (\ref{eq1})
is bounded above by
\begin{eqnarray*}
&& \bigl\|E_{\widehat{\mu}_{1}} \{\SGN_{{\mathbf x} - {\mathbf X}}\} - E_{\widehat{\mu
}_{1}}
\{\SGN_{{\mathbf v}_{j} - {\mathbf X}}\}\bigr\| + \bigl\|E_{\mu_{1}}\{\SGN_{{\mathbf x}
- {\mathbf X}}\} -
E_{\mu_{1}}\{\SGN_{{\mathbf v}_{j} - {\mathbf X}}\}\bigr\|
\\
&&\qquad {}+ \max_{1 \leq l \leq N(\varepsilon)} \bigl\|E_{\widehat{\mu}_{1}} \{ \SGN_{{\mathbf v}_{l} - {\mathbf X}}\} -
E_{\mu_{1}}\{\SGN_{{\mathbf v}_{l} - {\mathbf
X}}\}\bigr\|,
\end{eqnarray*}
where $\|{\mathbf x} - {\mathbf v}_{j}\| < \varepsilon$. Using Assumption \ref{assA}
in Section~\ref{sec2}, it follows that
%
%
\begin{eqnarray}\label{eq11}
\bigl\|E_{\widehat{\mu}_{1}} \{\SGN_{{\mathbf x} - {\mathbf X}}\} - E_{\widehat
{\mu}_{1}}
\{\SGN_{{\mathbf v}_{j} - {\mathbf X}}\}\bigr\| &\leq& E_{\widehat{\mu
}_{1}}\bigl\{T({\mathbf v}_{j} -
{\mathbf X})\bigr\}\|{\mathbf x} - {\mathbf v}_{j}\|
\nonumber\\[-8pt]\\[-8pt]
&\leq& 2{\varepsilon}E_{\mu_{1}}\bigl\{T({\mathbf v}_{j} - {\mathbf X})
\bigr\},\nonumber
\end{eqnarray}
for all $n$ sufficiently large \emph{almost surely}. Further,
%
%
\begin{eqnarray}
&& \bigl\|E_{\mu_{1}}\{\SGN_{{\mathbf x} - {\mathbf X}}\} - E_{\mu_{1}}\{\SGN_{{\mathbf
v}_{j} - {\mathbf X}}
\}\bigr\| \leq{\varepsilon}E_{\mu_{1}}\bigl\{T({\mathbf v}_{j} - {\mathbf X})
\bigr\}. \label{eq12}
\end{eqnarray}
Using (\ref{eq11}) and (\ref{eq12}), the moment condition in the
theorem and the fact that $\max_{1 \leq l \leq N(\varepsilon)} \|
E_{\widehat{\mu}_{1}} \{\SGN_{{\mathbf v}_{l} - {\mathbf X}}\} - E_{\mu_{1}}\{
\SGN_{{\mathbf v}_{l} - {\mathbf X}}\}\|$ converges to zero as $n \rightarrow
\infty$ \emph{almost surely}, we get the proof of part~(a) of the theorem.

(b) As argued in the proof of Theorem~\ref{teo}, it is enough to
consider the case $d = 1$. Using Theorems 1.5.4 and 1.5.7 in \cite
{VW96}, it follows that we only need to prove the asymptotic
equicontinuity \emph{in probability} of $\widehat{{\mathbf S}}_{{\mathbf g}}$
with respect to the norm in ${\mathcal X}$. Further, since $\mu$ is
assumed to be nonatomic, the map ${\mathbf x} \mapsto{\mathbf g}(\sqrt
{n}(\widehat{S}_{{\mathbf x}} - S_{{\mathbf x}}))$ is \emph{almost surely} $\mu
$-continuous. Since $K$ is compact, it follows that the process
$\widehat{{\mathbf S}}_{{\mathbf g}}$ is separable (see page~115 in \cite{VW96}).
Thus, in view of Corollary 2.2.8 in \cite{VW96} and the assumption of
the finiteness of the integral $\int_{0}^{1} \sqrt{\ln N(\varepsilon,K)}$ for each $\varepsilon> 0$, we will have the asymptotic
equicontinuity \emph{in probability} of $\widehat{{\mathbf S}}_{{\mathbf g}}$ if
we can show the sub-Gaussianity of the process (see page~101 in \cite
{VW96}) with respect to the metric induced by the norm in ${\mathcal
X}$. Since ${\mathbf g} \in{ \operatorname{cal} X}^{**}$, the empirical process $\widehat{{\mathbf S}}_{{\mathbf g}} = \{
\sqrt{n}[n^{-1} \sum_{i=1}^{n} {\mathbf g}(\SGN_{{\mathbf x} - {\mathbf X}_{i}}) -
E\{{\mathbf g}(\SGN_{{\mathbf x} - {\mathbf X}_{i}})\}]\dvtx  {\mathbf x} \in K\}$. Using
the Bernstein inequality for real-valued random variables and the
assumptions in the theorem, we have
\begin{eqnarray*}
&& P\bigl(\bigl|\widehat{{\mathbf S}}_{{\mathbf g}}({\mathbf x}) - \widehat{{\mathbf
S}}_{{\mathbf
g}}({\mathbf y})\bigr| > t\bigr) \leq2\exp \bigl\{-t^{2}/a_{1}
\|{\mathbf x} - {\mathbf y}\| ^{2} \bigr\}\qquad \forall n
\end{eqnarray*}
for a suitable constant $a_{1} > 0$. This proves the sub-Gaussianity of
the process and completes the proof of the first statement in part~(b)
of the theorem.

 For proving the second statement in part~(b) of the theorem, we
will need Fact~\ref{fact1} stated earlier. Using this, we have
\begin{eqnarray*}
P\bigl(\bigl|\widehat{{\mathbf S}}_{{\mathbf g}}({\mathbf x}) - \widehat{{\mathbf
S}}_{{\mathbf
g}}({\mathbf y})\bigr| > t\bigr) &\leq& P\bigl(\sqrt{n}\bigl\|(
\widehat{S}_{{\mathbf x}} - \widehat {S}_{{\mathbf y}}) - (S_{{\mathbf x}} -
S_{{\mathbf y}})\bigr\| > t\bigr)
\\
&\leq& 2\exp \bigl\{-t^{2}/a_{2}\|{\mathbf x} - {\mathbf y}
\|^{2} \bigr\}\qquad \forall n
\end{eqnarray*}
for an appropriate constant $a_{2} > 0$. This proves the
sub-Gaussianity of the process, and hence its weak convergence to a
tight stochastic process.
\end{pf*}

\begin{pf*}{Proof of Theorem~\ref{teo1}}
Since ${\mathcal X}$ is strictly convex, and $\mu$ is not completely
supported on a straight line in ${\mathcal X}$, the map ${\mathbf x}
\mapsto E\{\|{\mathbf x} - {\mathbf X}\| - \|{\mathbf X}\|\}$ is strictly convex.
Thus, using Exercise 4.2.12 in \cite{BV10}, we have the strict
monotonicity of the spatial distribution map. Let $\widetilde{g}({\mathbf
y},{\mathbf v}) = E\{\|{\mathbf y} - {\mathbf X}\| - \|{\mathbf X}\|\} - {\mathbf v}({\mathbf
y})$, where ${\mathbf y} \in{\mathcal X}$ and ${\mathbf v} \in{\mathcal
B}^{*}({\mathbf0},1)$. Since ${\mathcal X}$ is reflexive, it follows from
Remark 3.5 in \cite{Kemp87} that there exists a minimizer of $\widetilde
{g}$ in ${\mathcal X}$. Let us denote it by ${\mathbf x}({\mathbf v})$. So,
$\widetilde{g}({\mathbf x}({\mathbf v}),{\mathbf v}) \leq\widetilde{g}({\mathbf
y},{\mathbf v})$ for all ${\mathbf y} \in{\mathcal X}$. Equivalently, ${\mathbf
v}\{{\mathbf y} - {\mathbf x}({\mathbf v})\} \leq E\{\|{\mathbf y} - {\mathbf X}\| - \|
{\mathbf x}({\mathbf v}) - {\mathbf X}\|\}$ for all ${\mathbf y} \in{\mathcal X}$.
Since $\mu$ is nonatomic, it follows that the map ${\mathbf x} \mapsto E\{\|
{\mathbf x} - {\mathbf X}\| - \|{\mathbf
X}\|\}$ is G\^ateaux differentiable everywhere. So, using the previous
inequality and Corollary 4.2.5 in~\cite{BV10}, we have $S_{{\mathbf x}({\mathbf
v})} = E\{\SGN_{{\mathbf x}({\mathbf v}) - {\mathbf X}}\} = {\mathbf v}$. This proves
that the range of the spatial distribution map is the whole of
${\mathcal B}^{*}({\mathbf0},1)$. Since the norm in ${\mathcal X}$ is Fr\'
echet differentiable on ${\mathcal X}\setminus\{{\mathbf0}\}$ and $\mu$
is nonatomic, the map ${\mathbf x} \mapsto E\{\|{\mathbf x} - {\mathbf X}\| - \|
{\mathbf X}\|\}$ is Fr\'echet differentiable everywhere. The continuity
property of the spatial distribution map is now a consequence of
Corollary 4.2.12 in \cite{BV10}.
\end{pf*}
The next result can be obtained by suitably modifying the arguments in
the second paragraph in the proof of Theorem 3.1.1 in \cite{Chau96}.
%
\begin{fact} \label{lemma1}
If ${\mathcal X}$ is a Banach space, there exists $C_{1} > 0$
(depending on~${\mathbf u}$) such that $\|\widehat{{\mathbf Q}}({\mathbf u}) - {\mathbf
Q}({\mathbf u})\| \leq C_{1}$ for all sufficiently large $n$ \emph{almost surely}.
\end{fact}
\begin{pf*}{Proof of Theorem~\ref{teo5}}
From the assumptions in the theorem and Theorem~2.17 and Remark 3.5 in
\cite{Kemp87}, it follows that ${\mathbf Q}({\mathbf u})$ exists and is unique.
Let $\widehat{g}_{n}({\mathbf Q}) = n^{-1} \sum_{i=1}^{n} \{\|{\mathbf Q} -
{\mathbf X}_{i}^{(n)}\| - \|{\mathbf X}_{i}^{(n)}\|\} - {\mathbf u}^{(n)}({\mathbf Q})$
for ${\mathbf Q} \in{\mathcal X}$. We will first prove the result when
${\mathbf X}$ is assumed to be bounded \emph{almost surely}, that is, for
some $M > 0$, $P(\|{\mathbf X}\| \leq M) = 1$. Now, it follows from
arguments similar to those in the proof of Lemma~2(i) in \cite{Cadr01}
that $\sup_{\|{\mathbf Q}\| \leq C} |\widehat{g}_{n}({\mathbf Q}) - g_{n}({\mathbf
Q})| \rightarrow0$ as $n \rightarrow\infty$ \emph{almost surely} for
any $C > 0$. We next show that $g(\widehat{{\mathbf Q}}({\mathbf u}))
\rightarrow g({\mathbf Q}({\mathbf u}))$ as $n \rightarrow\infty$ \emph{almost
surely}. Note that
%
%
\begin{eqnarray}\label{eq50}
0 &\leq& g\bigl(\widehat{{\mathbf Q}}({\mathbf u})\bigr) - g\bigl({\mathbf Q}({\mathbf u})
\bigr)\nonumber
\\
& = &\bigl[g\bigl(\widehat{{\mathbf Q}}({\mathbf u})\bigr) - g_{n}\bigl(
\widehat{{\mathbf Q}}({\mathbf u})\bigr)\bigr]
- \bigl[g\bigl({\mathbf Q}({\mathbf u})\bigr) - g_{n}\bigl({
\mathbf Q}({\mathbf u})\bigr)\bigr]
\\
&&{} + \bigl[g_{n}\bigl(\widehat{{\mathbf Q}}({\mathbf
u})\bigr) - g_{n}\bigl({\mathbf Q}({\mathbf u})\bigr)\bigr].
\nonumber
\end{eqnarray}
Observe that for any ${\mathbf Q}$, $|g({\mathbf Q}) - g_{n}({\mathbf Q})| \leq2E\{
\|{\mathbf X} - {\mathbf X}^{(n)}\|\} + \|{\mathbf Q}\|\|{\mathbf u} - {\mathbf u}^{(n)}\|
$, which implies that
%
%
\begin{eqnarray}
&& \sup_{\|{\mathbf Q}\| \leq C} \bigl|g({\mathbf Q}) - g_{n}({\mathbf Q})\bigr|
\rightarrow0, \label{eq51}
\end{eqnarray}
as $n \rightarrow\infty$ \emph{almost surely} for any $C > 0$. Further,
%
%
\begin{eqnarray}\label{eqp51}
&& g_{n}\bigl(\widehat{{\mathbf Q}}({\mathbf u})\bigr) - g_{n}
\bigl({\mathbf Q}({\mathbf u})\bigr) \nonumber
\\
&&\qquad = \bigl[g_{n}\bigl(\widehat{{\mathbf Q}}({\mathbf u})\bigr) -
\widehat{g}_{n}\bigl(\widehat{{\mathbf Q}}({\mathbf u})\bigr)\bigr] + \bigl[
\widehat{g}_{n}\bigl(\widehat{{\mathbf Q}}({\mathbf u})\bigr) - \widehat
{g}_{n}\bigl({\mathbf Q}^{(n)}({\mathbf u})\bigr)\bigr]
\\
&&\quad\qquad{} + \bigl[\widehat{g}_{n}\bigl({\mathbf Q}^{(n)}({\mathbf u})\bigr)
- g_{n}\bigl({\mathbf Q}^{(n)}({\mathbf u})\bigr)\bigr] +
\bigl[g_{n}\bigl({\mathbf Q}^{(n)}({\mathbf u})\bigr) -
g_{n}\bigl({\mathbf Q}({\mathbf u})\bigr)\bigr].
\nonumber
\end{eqnarray}
In the notation of Section~\ref{sec4}, ${\mathbf Q}^{(n)}({\mathbf u}) = \sum_{k=1}^{d(n)} q_{k}\phi_{k}$, where ${\mathbf Q} = \sum_{k=1}^{\infty}
q_{k}\phi_{k}$ for a Schauder basis $\{\phi_{1},\phi_{2},\ldots\}$ of
${\mathcal X}$. The first and the third terms in the right-hand side of
(\ref{eqp51}) are bounded above by $\sup_{\|{\mathbf Q}\| \leq C_{2}}
|\widehat{g}_{n}({\mathbf Q}) - g_{n}({\mathbf Q})|$ for all sufficiently large
$n$ \emph{almost surely}. Here, $C_{2} = C_{1} + 2\|{\mathbf Q}({\mathbf u})\|$,
and $C_{1}$ is as in Fact~\ref{lemma1}. The second term in the
right-hand side of (\ref{eqp51}) is negative because $\widehat{{\mathbf
Q}}({\mathbf u})$ is a minimizer of $\widehat{g}_{n}$. The fourth term in
the right-hand side of (\ref{eqp51}) is bounded above by $2\|{\mathbf
Q}^{(n)}({\mathbf u}) - {\mathbf Q}({\mathbf u})\|$. So,
\[
g_{n}\bigl(\widehat{{\mathbf Q}}({\mathbf u})\bigr) - g_{n}\bigl({
\mathbf Q}({\mathbf u})\bigr) \leq2\sup_{\|{\mathbf Q}\| \leq C_{2}} \bigl|\widehat{g}_{n}({
\mathbf Q}) - g_{n}({\mathbf Q})\bigr| + 2\bigl\|{\mathbf Q}^{(n)}({\mathbf u}) - {\mathbf
Q}({\mathbf u})\bigr\|
\]
for\vspace*{2pt} all sufficiently large $n$ \emph{almost surely}. Combining (\ref
{eq50}), (\ref{eq51}) and the previous inequality, we get $g(\widehat
{{\mathbf Q}}({\mathbf u})) \rightarrow g({\mathbf Q}({\mathbf u}))$ as $n \rightarrow
\infty$ \emph{almost surely}.

 Let us now observe that for any random element ${\mathbf X}$ in the
separable Banach space ${\mathcal X}$ and any fixed $\varepsilon> 0$,
there exists $M > 0$ such that $P(\|{\mathbf X}\| > M) <\varepsilon/C_{1}$.
So, we have $|g(\widehat{{\mathbf Q}}({\mathbf u})) - g({\mathbf Q}({\mathbf u}))| \leq
\varepsilon+ |\overline{g}(\widehat{{\mathbf Q}}({\mathbf u})) - \overline
{g}({\mathbf Q}({\mathbf u}))|$ for all sufficiently large $n$ \emph{almost
surely}. Here, $\overline{g}({\mathbf Q}) = E\{(\|{\mathbf Q} - {\mathbf X}\| - \|
{\mathbf X}\|)I(\|{\mathbf X}\| \leq M)\} - {\mathbf u}({\mathbf Q})$. Thus, letting
$\varepsilon\rightarrow0$, we have $g(\widehat{{\mathbf Q}}({\mathbf u}))
\rightarrow g({\mathbf Q}({\mathbf u}))$ as $n \rightarrow\infty$ \emph{almost
surely} for those random elements in ${\mathcal X}$ that are not
necessarily \emph{almost surely} bounded. Now, using Theorems 1 and 3 in
\cite{Aspl68}, it follows that $\|\widehat{{\mathbf Q}}({\mathbf u}) - {\mathbf
Q}({\mathbf u})\| \rightarrow0$ as $n \rightarrow\infty$ \emph{almost surely}.
\end{pf*}

The Hessian of the function $g_{n}({\mathbf Q})$ is
\begin{eqnarray*}
J_{n,{\mathbf Q}}({\mathbf h},{\mathbf v}) &=& E \biggl\{ \frac{\langle{\mathbf h},{\mathbf
v}\rangle}{\|{\mathbf Q} - {\mathbf X}^{(n)}\|} -
\frac{\langle{\mathbf h},{\mathbf Q}
- {\mathbf X}^{(n)}\rangle\langle{\mathbf v},{\mathbf Q} - {\mathbf X}^{(n)}\rangle}{\|
{\mathbf Q} - {\mathbf X}^{(n)}\|^{3}} \biggr\}.
\end{eqnarray*}
The next result is the $d(n)$-dimensional analog of Proposition 2.1 in
\cite{CCZ13}, and can be obtained by suitably modifying the proof of
that proposition.
%
\begin{fact} \label{lemma2}
Suppose that the assumptions of Theorem~\ref{teo6} hold. Then, for each
$C > 0$, there exists $b, B \in(0,\infty)$ with $b < B$ such that for
all appropriately large $n$ we have $b\|{\mathbf h}\|^{2} \leq J_{n,{\mathbf
Q}}({\mathbf h},{\mathbf h}) \leq B\|{\mathbf h}\|^{2}$ for any ${\mathbf Q}$, ${\mathbf h}
\in{\mathcal Z}_{n}$ with $\|{\mathbf Q}\| \leq C$.
\end{fact}

%
\begin{lemma} \label{lemma3}
Suppose that the assumptions of Theorem~\ref{teo6} hold and $C > 0$ is
arbitrary. Then there exist $b', B'\in(0,\infty)$ such that for all
appropriately large $n$ and any ${\mathbf Q}$, ${\mathbf h}$, ${\mathbf z} \in
{\mathcal Z}_{n}$ with $\|{\mathbf Q} - {\mathbf Q}_{n}({\mathbf u})\| \leq C$, we have
\begin{eqnarray*}
\biggl\llVert E \biggl\{\frac{{\mathbf Q} - {\mathbf X}^{(n)}}{\|{\mathbf Q} - {\mathbf
X}^{(n)}\|} - {\mathbf u}^{(n)} \biggr\}
\biggr\rrVert &\geq& b'\bigl\|{\mathbf Q} - {\mathbf Q}_{n}({\mathbf u})\bigr\|,
\\
\sup_{\|{\mathbf h}\| = \|{\mathbf v}\| = 1} \bigl|J_{n,{\mathbf Q}}({\mathbf h},{\mathbf v}) -
J_{n,{\mathbf Q}_{n}({\mathbf u})}({\mathbf h},{\mathbf v})\bigr| &\leq& B'\bigl\|{\mathbf Q} - {\mathbf
Q}_{n}({\mathbf u})\bigr\|
\end{eqnarray*}
and
\[
\biggl\llVert E \biggl\{\frac{{\mathbf Q} - {\mathbf X}^{(n)}}{\|{\mathbf Q} - {\mathbf
X}^{(n)}\|} - {\mathbf u}^{(n)} \biggr\}
- \widetilde{J}_{n,{\mathbf Q}_{n}({\mathbf
u})}\bigl({\mathbf Q} - {\mathbf Q}_{n}({\mathbf u})
\bigr)\biggr\rrVert \leq B'\bigl\|{\mathbf Q} - {\mathbf Q}_{n}({\mathbf
u})\bigr\|^{2}.
\]
\end{lemma}

\begin{pf} For any $\|{\mathbf h}\| = 1$, a first order Taylor expansion of
the function $E \{\frac{{\mathbf Q} - {\mathbf X}^{(n)}}{\|{\mathbf Q} - {\mathbf
X}^{(n)}\|} - {\mathbf u}^{(n)} \}({\mathbf h})$ about ${\mathbf Q}_{n}({\mathbf
u})$ yields
%
%
\begin{equation}
E \biggl\{\frac{{\mathbf Q} - {\mathbf X}^{(n)}}{\|{\mathbf Q} - {\mathbf X}^{(n)}\|} - {\mathbf u}^{(n)} \biggr\}({\mathbf h}) =
J_{n,\widetilde{{\mathbf Q}}}\bigl({\mathbf Q} - {\mathbf Q}_{n}({\mathbf u}),{\mathbf h}\bigr),
\label{eq2}
\end{equation}
where $\|\widetilde{{\mathbf Q}} - {\mathbf Q}_{n}({\mathbf u})\| < \|{\mathbf Q} -
{\mathbf Q}_{n}({\mathbf u})\|$. Choosing ${\mathbf h} = ({\mathbf Q} - {\mathbf
Q}_{n}({\mathbf u}))/\|{\mathbf Q} - {\mathbf Q}_{n}({\mathbf u})\|$ and using Fact~\ref
{lemma2}, we have the first inequality.\vadjust{\goodbreak}

 The second inequality follows from the definition of $J_{n,Q}$,
the upper bound in Fact~\ref{lemma2} and some straight-forward algebra.

 From (\ref{eq2}), we get
\begin{eqnarray*}
&& \biggl\llvert E \biggl\{\frac{{\mathbf Q} - {\mathbf X}^{(n)}}{\|{\mathbf Q} - {\mathbf
X}^{(n)}\|} - {\mathbf u}^{(n)} \biggr
\}({\mathbf h}) - J_{n,{\mathbf Q}_{n}({\mathbf
u})}\bigl({\mathbf Q} - {\mathbf Q}_{n}({\mathbf u}),{
\mathbf h}\bigr)\biggr\rrvert
\\
&&\qquad = \bigl|J_{n,\widetilde{{\mathbf Q}}}\bigl({\mathbf Q} - {\mathbf Q}_{n}({\mathbf u}),{\mathbf h}
\bigr) - J_{n,{\mathbf Q}_{n}({\mathbf u})}\bigl({\mathbf Q} - {\mathbf Q}_{n}({\mathbf u}),{\mathbf h}
\bigr)\bigr|
\\
&&\qquad \leq B'\bigl\|{\mathbf Q} - {\mathbf Q}_{n}({\mathbf u})
\bigr\|^{2},\qquad\mbox{since $\bigl\| \widetilde{{\mathbf Q}} - {\mathbf Q}_{n}({
\mathbf u})\bigr\| < \bigl\|{\mathbf Q} - {\mathbf Q}_{n}({\mathbf u})\bigr\|$}.
\end{eqnarray*}
Taking supremum over $\|{\mathbf h}\| = 1$ and using the definition of
$\widetilde{J}_{n,{\mathbf Q}}$, we have the proof of the third inequality.
\end{pf}

%
\begin{proposition} \label{prop1}
Suppose that the assumptions of Theorem~\ref{teo6} hold. Then, $\|
\widehat{{\mathbf Q}}({\mathbf u}) - {\mathbf Q}_{n}({\mathbf u})\| = O(\delta_{n})$ as
n $\rightarrow\infty$ \emph{almost surely}, where $\delta_{n} \sim\sqrt
{\ln  n}/n^{\alpha}$ and $\alpha$ is as in Theorem~\ref{teo6}.
\end{proposition}
\begin{pf} From Fact~\ref{lemma1} and the behavior of ${\mathbf Q}_{n}({\mathbf
u})$ discussed before Assumption \ref{assB} in Section~\ref{subsec42}, we get
the existence of $C_{3} > 0$ satisfying $\|\widehat{{\mathbf Q}}({\mathbf u}) -
{\mathbf Q}_{n}({\mathbf u})\| \leq C_{3}$ for all sufficiently large $n$ \emph{almost surely}. Define $\mbox{G}_{n} = \{{\mathbf Q}_{n}({\mathbf u}) + \sum_{j
\leq d(n)} \beta_{j}\varphi_{j}\dvtx  n^{4}\beta_{j}$ is an integer in
$[-C_{3},C_{3}]$ and $\|\sum_{j \leq d(n)} \beta_{j}\varphi_{j}\| \leq
C_{3} \}$, and ${\mathcal Z}_{n} = \span\{\varphi_{1},\varphi_{2},\ldots,\varphi_{d(n)}\}$, where $\{\varphi_{j}\}_{j \geq1}$ is an
orthonormal basis of ${\mathcal X}$. Let us define the event
\begin{eqnarray*}
\mbox{E}_{n} &\,{=}\,& \Biggl\{ \max_{{\mathbf Q} \in{\rm{G}}_{n}} \Biggl\llVert
\frac
{1}{n}\sum_{i=1}^{n} \!\biggl(
\frac{{\mathbf Q} \,{-}\, {\mathbf X}_{i}^{(n)}}{\|{\mathbf
Q} \,{-}\, {\mathbf X}_{i}^{(n)}\|} \,{-}\, {\mathbf u}^{(n)}\! \biggr) \,{-}\,
E \biggl( \frac{{\mathbf Q}\,{-}\, {\mathbf X}^{(n)}}{\|{\mathbf Q} \,{-}\, {\mathbf
X}^{(n)}\|}\,{-}\, {\mathbf u}^{(n)} \biggr)\! \Biggr\rrVert \,{\leq}\, C_{4}
\delta_{n} \Biggr\}.
\end{eqnarray*}
Note that $\llVert \frac{{\mathbf Q} - {\mathbf X}^{(n)}}{\|{\mathbf Q} - {\mathbf
X}^{(n)}\|} - {\mathbf u}^{(n)}\rrVert  \leq2$ for all ${\mathbf Q} \in
{\mathcal Z}_{n}$ and $n \geq1$. So, using Fact~\ref{fact1}, there
exists $C_{5} > 0$ such that $P(\mbox{E}_{n}^{c}) \leq
2(3C_{3}n^{4})^{d(n)}\exp\{-nC_{5}^{2}\delta^{2}_{n}\}$ for all
appropriately large $n$. Using the definition of $\delta_{n}$ given in
the statement of the proposition, $C_{5}$ in the previous inequality
can be chosen in such a way that $\sum_{n=1}^{\infty} P(\mbox
{E}_{n}^{c}) < \infty$. Thus,
%
%
\begin{equation}
P(\mbox{E}_{n} \mbox{ occurs for all sufficiently large } n) = 1.
\label{eq3}
\end{equation}
We\vspace*{-1pt} next define the event $\mbox{F}_{n} =  \{\max_{{\mathbf Q} \in{\rm
{G}}_{n}} \sum_{i=1}^{n}
I_{\{\|{\mathbf Q} - {\mathbf X}_{i}^{(n)}\| \leq n^{-2}\}} \leq C_{6}n\delta
^{2}_{n} \}$. Note
that $M'_{n} = \max_{{\mathbf Q} \in{\rm{G}}_{n}} E\{\|{\mathbf Q} - {\mathbf
X}^{(n)}\|^{-1}\} < \infty$
for all appropriately large $n$ in view of Assumption \ref{assB} in
Section~\ref{subsec42}. Further,
$M'_{n} \geq M'_{n+k}$ for all $k \geq1$ and $n \geq1$. Then, $P(\|
{\mathbf Q} - {\mathbf X}^{(n)}\|
\leq n^{-2}) \leq M'_{n}n^{-2} \leq C_{6}\delta^{2}_{n}/2$ for any
${\mathbf Q} \in\mbox{G}_{n}$
and all appropriately large $n$ (the first inequality follows from the
Markov inequality).
Therefore, $\Var\{I(\|{\mathbf Q} - {\mathbf X}^{(n)}\| \leq n^{-2})\} \leq
C_{6}\delta^{2}_{n}/2$ for
any ${\mathbf Q} \in\mbox{G}_{n}$ and all appropriately large~$n$. The
Bernstein inequality for
real-valued random variables implies that there exists $C_{7} > 0$ such that
$P(\mbox{F}_{n}^{c}) \leq(3C_{3}n^{4})^{d(n)}\exp\{-nC_{7}\delta
^{2}_{n}\}$ for all
appropriately large $n$. As before, $C_{7}$ in the previous inequality
can be chosen in such a
way that $\sum_{n=1}^{\infty} P(\mbox{F}_{n}^{c}) < \infty$, which
implies that
%
%
\begin{equation}
P(\mbox{F}_{n} \mbox{ occurs for all sufficiently large } n) = 1.
\label{eq4}
\end{equation}
Now consider a point in $\mbox{G}_{n}$ nearest to $\widehat{{\mathbf
Q}}({\mathbf u})$, say, $\overline{{\mathbf Q}}_{n}({\mathbf u})$. Then, $\|\widehat
{{\mathbf Q}}({\mathbf u}) - \overline{{\mathbf Q}}_{n}({\mathbf u})\| \leq
C_{8}\,d(n)/n^{4}$ for a constant $C_{8} > 0$. Note that
%
%
\begin{eqnarray}
&& \biggl\llVert \frac{\widehat{{\mathbf Q}}({\mathbf u}) - {\mathbf X}_{i}^{(n)}}{\|
\widehat{{\mathbf Q}}({\mathbf u}) - {\mathbf X}_{i}^{(n)}\|} - \frac{\overline
{{\mathbf Q}}_{n}({\mathbf u}) - {\mathbf X}_{i}^{(n)}}{\|\overline{{\mathbf
Q}}_{n}({\mathbf u}) - {\mathbf X}_{i}^{(n)}\|}\biggr\rrVert \leq
\frac{2\|\widehat{{\mathbf Q}}({\mathbf u}) - \overline{{\mathbf Q}}_{n}({\mathbf u})\|
}{\|\overline{{\mathbf Q}}_{n}({\mathbf u}) - {\mathbf X}_{i}^{(n)}\|}. \label{eq5}
\end{eqnarray}
Then, for a constant $C_{9} > 0$, we have
%
%
\begin{eqnarray}\label{eq6}
&& \Biggl\llVert \frac{1}{n} \sum_{i=1}^{n}
\frac{\overline{{\mathbf Q}}_{n}({\mathbf
u}) - {\mathbf
X}_{i}^{(n)}}{\|\overline{{\mathbf Q}}_{n}({\mathbf u}) - {\mathbf X}_{i}^{(n)}\|} - {\mathbf u}^{(n)} \Biggr\rrVert\nonumber
\\
&&\qquad \leq\Biggl\llVert
\frac{1}{n} \sum_{i=1}^{n}
\frac{\widehat{{\mathbf Q}}({\mathbf
u}) - {\mathbf X}_{i}^{(n)}}{\|\widehat{{\mathbf Q}}({\mathbf u}) - {\mathbf X}_{i}^{(n)}\|} - {\mathbf u}^{(n)} \Biggr\rrVert \nonumber
\\
&&\quad\qquad{} + \Biggl\llVert \frac{1}{n} \sum_{i=1}^{n}
\biggl\{ \frac{\overline{{\mathbf
Q}}_{n}({\mathbf u}) - {\mathbf
X}_{i}^{(n)}}{\|\overline{{\mathbf Q}}_{n}({\mathbf u}) - {\mathbf X}_{i}^{(n)}\|} - \frac{\widehat{{\mathbf
Q}}({\mathbf u}) - {\mathbf X}_{i}^{(n)}}{\|\widehat{{\mathbf Q}}({\mathbf u}) - {\mathbf
X}_{i}^{(n)}\|} \biggr\} \Biggr\rrVert
\nonumber\\[-8pt]\\[-8pt]
&&\qquad \leq \Biggl\llVert \frac{1}{n} \sum_{i=1}^{n}
\frac{\widehat{{\mathbf Q}}({\mathbf
u}) - {\mathbf X}_{i}^{(n)}}{\|\widehat{{\mathbf Q}}({\mathbf u}) - {\mathbf
X}_{i}^{(n)}\|} - {\mathbf u}^{(n)} \Biggr\rrVert + 2C_{8}\,d(n)n^{-2}
\nonumber
\\
&&\quad\qquad{} + \frac{2}{n} \sum_{i=1}^{n} I
\bigl\{\bigl\|\overline{{\mathbf Q}}_{n}({\mathbf u}) - {\mathbf X}_{i}^{(n)}
\bigr\| \leq n^{-2}\bigr\}\qquad \bigl(\mbox{using (\ref{eq5})}\bigr)
\nonumber
\\
&&\qquad \leq \Biggl\llVert \frac{1}{n} \sum_{i=1}^{n}
\frac{\widehat{{\mathbf Q}}({\mathbf
u}) - {\mathbf X}_{i}^{(n)}}{\|\widehat{{\mathbf Q}}({\mathbf u}) - {\mathbf
X}_{i}^{(n)}\|} - {\mathbf u}^{(n)} \Biggr\rrVert + C_{9}
\delta^{2}_{n}\qquad \bigl(\mbox {using (\ref{eq4})}\bigr).
\nonumber
\end{eqnarray}
It follows from arguments similar to those used in the proof of Theorem
4.11 in~\cite{Kemp87} that $\llVert \sum_{i=1}^{n} \frac{\widehat{{\mathbf
Q}}({\mathbf u}) - {\mathbf X}_{i}^{(n)}}{\|\widehat{{\mathbf Q}}({\mathbf u}) - {\mathbf
X}_{i}^{(n)}\|} - n{\mathbf u}^{(n)}\rrVert  \leq1$. Combining this with
(\ref{eq6}), we get
%
%
\begin{equation}
\Biggl\llVert \sum_{i=1}^{n}
\frac{\overline{{\mathbf Q}}_{n}({\mathbf u}) - {\mathbf
X}_{i}^{(n)}}{\|\overline{{\mathbf Q}}_{n}({\mathbf u}) - {\mathbf X}_{i}^{(n)}\|} - n{\mathbf u}^{(n)}\Biggr\rrVert \leq3C_{7}n
\delta_{n} \label{eq61}
\end{equation}
for all sufficiently large $n$ \emph{almost surely}. Suppose that ${\mathbf
Q} \in\mbox{G}_{n}$ and $\|{\mathbf Q} - {\mathbf Q}_{n}({\mathbf u})\| >
C_{10}\delta_{n}$ for some $C_{10} > 0$. Then, it follows from (\ref
{eq3}) and the first inequality in Lemma~\ref{lemma3} that $\llVert \sum_{i=1}^{n} \frac{{\mathbf Q} - {\mathbf X}_{i}^{(n)}}{\|{\mathbf Q} - {\mathbf
X}_{i}^{(n)}\|} - n{\mathbf u}^{(n)}\rrVert  \geq(C_{10}b'-C_{4})n\delta
_{n}$ for all sufficiently large $n$ \emph{almost surely}. If we choose
$C_{10}$ such that $C_{10}b' - C_{4} > 4C_{7}$, then in view of (\ref
{eq61}), we must have $\|\overline{{\mathbf Q}}_{n}({\mathbf u}) - {\mathbf
Q}_{n}({\mathbf u})\| \leq C_{10}\delta_{n}$ for all sufficiently large $n$
\emph{almost surely}. This implies that for a constant $C_{11} > 0$, $\|
\widehat{{\mathbf Q}}({\mathbf u}) - {\mathbf Q}_{n}({\mathbf u})\| \leq C_{11}\delta
_{n}$ for all sufficiently large $n$ \emph{almost surely}. This
completes the proof.
\end{pf}

\begin{pf*}{Proof of Theorem~\ref{teo6}}
Let $\mbox{H}_{n}$ denote the collection of points from $\mbox{G}_{n}$,
which satisfy $\|{\mathbf Q} - {\mathbf Q}_{n}({\mathbf u})\| \leq C_{11}\delta
_{n}$. Let us define for ${\mathbf Q} \in{\mathcal Z}_{n}$,
\[
\Gamma_{n}({\mathbf Q},{\mathbf X}_{i}) = \frac{{\mathbf Q}_{n}({\mathbf u}) - {\mathbf
X}_{i}^{(n)}}{\|{\mathbf Q}_{n}({\mathbf u}) - {\mathbf X}_{i}^{(n)}\|} -
\frac
{{\mathbf Q} - {\mathbf X}_{i}^{(n)}}{\|{\mathbf Q} - {\mathbf X}_{i}^{(n)}\|} + E \biggl\{ \frac{{\mathbf Q} - {\mathbf X}^{(n)}}{\|{\mathbf Q} - {\mathbf X}^{(n)}\|} - {\mathbf u}^{(n)} \biggr
\}
\]
and
\[
\Delta_{n}({\mathbf Q}) = E \biggl\{ \frac{{\mathbf Q} - {\mathbf
X}^{(n)}}{\|{\mathbf Q} - {\mathbf X}^{(n)}\|} -
\frac{{\mathbf Q}_{n}({\mathbf u}) -
{\mathbf X}^{(n)}}{\|{\mathbf Q}_{n}({\mathbf u}) - {\mathbf X}^{(n)}\|} \biggr\} - \widetilde{J}_{n,{\mathbf Q}_{n}({\mathbf u})}\bigl({\mathbf Q} - {\mathbf
Q}_{n}({\mathbf u})\bigr).
\]
Using Assumption \ref{assB} in Section~\ref{subsec42}, it follows that for a
constant $C_{12} > 0$,
\begin{eqnarray*}
E\bigl\llVert \Gamma_{n}({\mathbf Q},{\mathbf X}) \bigr\rrVert
^{2} &\leq& 2E\biggl\llVert \frac
{{\mathbf Q}_{n}({\mathbf u}) -
{\mathbf X}_{i}^{(n)}}{\|{\mathbf Q}_{n}({\mathbf u}) - {\mathbf X}_{i}^{(n)}\|} - \frac
{{\mathbf Q} - {\mathbf
X}_{i}^{(n)}}{\|{\mathbf Q} - {\mathbf X}_{i}^{(n)}\|}
\biggr\rrVert ^{2}
\\
&&{} +
2\biggl\llVert E \biggl\{ \frac{{\mathbf Q}_{n}({\mathbf u}) - {\mathbf X}^{(n)}}{\|
{\mathbf Q}_{n}({\mathbf u}) -
{\mathbf X}^{(n)}\|} \biggr\} - E \biggl\{
\frac{{\mathbf Q} - {\mathbf X}^{(n)}}{\|
{\mathbf Q} - {\mathbf
X}^{(n)}\|} \biggr\}\biggr\rrVert ^{2}
\\
&&\qquad \leq C_{12}\bigl\|{\mathbf Q} - {\mathbf Q}_{n}({\mathbf u})
\bigr\|^{2}.
\end{eqnarray*}
So, in view of Fact~\ref{fact1}, there exists a constant $C_{13} > 0$
such that
%
%
\begin{equation}
\max_{{\mathbf Q} \in{\rm{H}}_{n}} \Biggl\llVert \frac{1}{n} \sum
_{i=1}^{n} \Gamma_{n}({\mathbf Q},{\mathbf
X}_{i}) \Biggr\rrVert \leq C_{13}\delta^{2}_{n},
\label{eq7}
\end{equation}
for all sufficiently large $n$ \emph{almost surely}. Using the third
inequality in Lem\-ma~\ref{lemma3}, there exists a constant $C_{14} > 0$
such that $\|\Delta_{n}({\mathbf Q})\| \leq C_{14}\|{\mathbf Q} - {\mathbf
Q}_{n}({\mathbf u})\|^{2}$ for all appropriately large $n$. This along with
(\ref{eq7}) and the definitions of $\Gamma_{n}$ and $\Delta_{n}({\mathbf
Q})$ yield
\[
\widetilde{J}_{n,{\mathbf Q}_{n}({\mathbf u})}\bigl({\mathbf Q} - {\mathbf Q}_{n}({\mathbf u})
\bigr) = \frac{1}{n} \sum_{i=1}^{n}
\biggl\{ \frac{{\mathbf Q}_{n}({\mathbf u}) - {\mathbf
X}_{i}^{(n)}}{\|{\mathbf Q}_{n}({\mathbf u}) - {\mathbf X}_{i}^{(n)}\|} - \frac
{{\mathbf Q} - {\mathbf X}_{i}^{(n)}}{\|{\mathbf Q} - {\mathbf X}_{i}^{(n)}\|} \biggr\} + \widetilde{{\mathbf
R}}_{n}({\mathbf Q}),
\]
where\vspace*{1pt} $\max_{{\mathbf Q} \in{\rm{H}}_{n}} \|\widetilde{{\mathbf R}}_{n}({\mathbf
Q})\| = O(\delta^{2}_{n})$ as $n \rightarrow\infty$ \emph{almost
surely}. From Fact~\ref{lemma2}, it follows that the operator norm of
$\widetilde{J}_{n,{\mathbf Q}_{n}({\mathbf u})}$ is uniformly bounded away from
zero, and $[\widetilde{J}_{n,{\mathbf Q}_{n}({\mathbf u})}]^{-1}$ is defined on
the whole of ${\mathcal Z}_{n}$ for all appropriately large $n$. It
follows that for a constant $C_{15} > 0$, $\max_{{\mathbf Q} \in{\rm
{H}}_{n}}\|[\widetilde{J}_{n,{\mathbf Q}_{n}({\mathbf u})}]^{-1}(\widetilde
{{\mathbf R}}_{n}({\mathbf Q}))\| \leq C_{15}\delta^{2}_{n}$ for all
sufficiently large $n$ \emph{almost surely}.

 Hence, choosing ${\mathbf Q} = \overline{{\mathbf Q}}_{n}({\mathbf u})$,
and utilizing inequality (\ref{eq6}) in the proof of Proposition~\ref
{prop1}, we get
\[
\widehat{{\mathbf Q}}({\mathbf u}) - {\mathbf Q}_{n}({\mathbf u}) = \frac{1}{n}
\sum_{i=1}^{n} [\widetilde{J}_{n,{\mathbf Q}_{n}({\mathbf u})}]^{-1}
\biggl\{\frac
{{\mathbf Q}_{n}({\mathbf u}) - {\mathbf X}_{i}^{(n)}}{\|{\mathbf Q}_{n}({\mathbf u}) -
{\mathbf X}_{i}^{(n)}\|} - {\mathbf u}^{(n)} \biggr\} + {\mathbf
R}_{n},
\]
where $\|{\mathbf R}_{n}\| = O(\delta^{2}_{n})$ as $n \rightarrow\infty$
\emph{almost surely}.
\end{pf*}

\begin{pf*}{Proof of Theorem~\ref{teo7}}
Since ${\mathbf U}_{n} = n^{-1}\sum_{i=1}^{n}  (\frac{{\mathbf Q}_{n}({\mathbf
u}) - {\mathbf X}_{i}^{(n)}}{\|{\mathbf Q}_{n}({\mathbf u}) - {\mathbf X}_{i}^{(n)}\|}
- {\mathbf u}^{(n)} )$ is a sum of uniformly bounded, independent,
zero mean random elements in the separable Hilbert space ${\mathcal
X}$, we get that $\|\sqrt{n}{\mathbf U}_{n}\|$ is bounded\vspace*{1pt} \emph{in
probability} as $n \rightarrow\infty$ in view of Fact~\ref{fact1}. We
will show that $\sqrt{n}\{[\widetilde{J}_{n,{\mathbf Q}_{n}({\mathbf
u})}]^{-1}({\mathbf U}_{n}) - [\widetilde{J}_{{\mathbf Q}({\mathbf u})}]^{-1}({\mathbf
U}_{n})\} \rightarrow{\mathbf0}$ \emph{in probability} as $n \rightarrow
\infty$. Note that for each $C > 0$, every\vspace*{1pt} ${\mathbf Q} \in{\mathcal X}$
satisfying \mbox{$\|{\mathbf Q}\| \leq C$} and all appropriately large $n$,
$J_{n,{\mathbf Q}}$ and $\widetilde{J}_{n,{\mathbf Q}}$ can be defined from
${\mathcal X} \times{\mathcal X} \rightarrow\mathbb{R}$ and
${\mathcal X} \rightarrow{\mathcal X}$, respectively, by virtue of
Assumption \ref{assB} in Section~\ref{subsec42}. Further, the bound obtained
in the second inequality in
Lemma~\ref{lemma3} actually holds (up to a constant multiple) for all\vspace*{2pt}
appropriately large $n$, any $C > 0$ and any ${\mathbf Q}$, ${\mathbf h}$,
${\mathbf v} \in{\mathcal X}$, which satisfy $\|{\mathbf Q}\| \leq C$. Thus,
$\|\widetilde{J}_{n,{\mathbf Q}_{n}({\mathbf u})} - \widetilde{J}_{n,{\mathbf
Q}({\mathbf u})}\| \leq B''\|{\mathbf Q}_{n}({\mathbf u}) - {\mathbf Q}({\mathbf u})\|$ for\vspace*{1pt}
a constant $B'' > 0$ and all appropriately large $n$. Since $\|{\mathbf
X}^{(n)} - {\mathbf X}\| \rightarrow0$ as $n \rightarrow\infty$ \emph{almost surely}, it follows from Assumption \ref{assB} in Section~\ref
{subsec42} that $\|\widetilde{J}_{n,{\mathbf Q}({\mathbf u})} - \widetilde
{J}_{{\mathbf Q}({\mathbf u})}\| \rightarrow0$ as $n \rightarrow\infty$.\vspace*{3pt}
Since ${\mathbf Q}_{n}({\mathbf u}) \rightarrow{\mathbf Q}({\mathbf u})$, we now have
$\|\widetilde{J}_{n,{\mathbf Q}_{n}({\mathbf u})} - \widetilde{J}_{{\mathbf Q}({\mathbf
u})}\| \rightarrow0$ as $n \rightarrow\infty$. It\vspace*{2pt} follows from
Proposition 2.1 in \cite{CCZ13} that the linear operator $\widetilde
{J}_{{\mathbf Q}({\mathbf u})}$ has a bounded inverse, which is defined on the
whole of~${\mathcal X}$. Using the fact that $\|\sqrt{n}{\mathbf U}_{n}\|$
is bounded \emph{in probability} as $n \rightarrow\infty$ we get that
\begin{eqnarray*}
&& \sqrt{n}\bigl\|\{\widetilde{J}_{n,{\mathbf Q}_{n}({\mathbf u})}\}^{-1}({\mathbf
U}_{n}) - \{\widetilde{J}_{{\mathbf Q}({\mathbf u})}\}^{-1}({\mathbf
U}_{n})\bigr\|
\\
&&\qquad \leq \sqrt{n}\bigl\|\{\widetilde{J}_{n,{\mathbf Q}_{n}({\mathbf u})}\}^{-1} - \{
\widetilde{J}_{{\mathbf Q}({\mathbf u})}\}^{-1}\bigr\|\bigl\|({\mathbf U}_{n})\bigr\|
\\
&&\qquad \leq \bigl\|\{\widetilde{J}_{{\mathbf Q}({\mathbf u})}\}^{-1}\bigr\|
\|\widetilde {J}_{n,{\mathbf Q}_{n}({\mathbf u})} - \widetilde{J}_{{\mathbf Q}({\mathbf u})}\|
\bigl\|\{ \widetilde{J}_{n,{\mathbf Q}_{n}({\mathbf u})}
\}^{-1}\bigr\|\|\sqrt{n} {\mathbf U}_{n}\|
\\
&&\qquad \stackrel{P} {\rightarrow} 0\qquad\mbox{as } n \rightarrow\infty.
\end{eqnarray*}
 The convergence \emph{in probability} asserted above holds
because the operator norm of $\widetilde{J}_{n,{\mathbf Q}_{n}({\mathbf u})}$
is uniformly bounded away from zero by Fact~\ref{lemma2}. The
asymptotic Gaussianity of $\{\widetilde{J}_{{\mathbf Q}({\mathbf u})}\}
^{-1}(\sqrt{n}{\mathbf U}_{n})$ follows from the central limit theorem for
a triangular array of row-wise independent Hilbert space valued random
elements (see, e.g., Corollary 7.8 in \cite{AG80}).
\end{pf*}

\begin{pf*}{Proof of Theorem~\ref{teo4}}
The proof of the first statement follows directly from part~(a) of
Theorem~\ref{teo2} after using the inequality $|\|{\mathbf x}\| - \|{\mathbf
y} \|| \leq\|{\mathbf x} - {\mathbf y}\|$, which holds for any ${\mathbf x},{\mathbf
y} \in{\mathcal X}$.

 Let us next consider the case $S_{{\mathbf x}} \neq{\mathbf0}$. From
the Fr\'echet differentiability of the norm in ${\mathcal X}^{*}$, we
have $\widehat{\SD}({\mathbf x}) - \SD({\mathbf x}) = \SGN_{S_{{\mathbf x}}}(\widehat
{S}_{{\mathbf x}} - S_{{\mathbf x}}) + o(\|\widehat{S}_{{\mathbf x}} - S_{{\mathbf x}}\|
)$. The central limit theorem for i.i.d. random elements in ${\mathcal
X}^{*}$ (see, e.g.,~\cite{AG80}) implies that $\sqrt{n}(\widehat
{S}_{{\mathbf x}} - S_{{\mathbf x}})$ converges\vspace*{2pt} weakly to a zero mean Gaussian
random element ${\mathbf W} \in{\mathcal X}^{*}$ as $n \rightarrow\infty
$. In particular, $\sqrt{n}\|\widehat{S}_{{\mathbf x}} - S_{{\mathbf x}}\|$ is
bounded \emph{in probability} as $n \rightarrow\infty$. Since the map
$\SGN_{S_{{\mathbf x}}}\dvtx  {\mathcal X}^{*} \rightarrow\mathbb{R}$ is
continuous, we now have the result for $S_{{\mathbf x}} \neq{\mathbf0}$ using
the continuous mapping theorem.

 Now, we consider the case $S_{{\mathbf x}} = {\mathbf0}$. In this
case, $\widehat{\SD}({\mathbf x}) - \SD({\mathbf x}) = - \|\widehat{S}_{{\mathbf x}}\|
$. The central limit theorem for i.i.d. random elements in ${\mathcal
X}^{*}$ yields that $\sqrt{n}\widehat{S}_{{\mathbf x}}$ converges weakly to
a zero mean Gaussian random element ${\mathbf V} \in{\mathcal X}^{*}$ as
$n \rightarrow\infty$. Finally, the continuous mapping theorem
completes the proof in view of the continuity of the norm function in
any Banach space.
\end{pf*}
\end{appendix}

\section*{Acknowledgements}
We thank an Associate Editor and a referee for their helpful comments.



\printaddresses
\end{document}